\documentclass[prb,aps,twocolumn,showpacs,amsmath,floatfix,superscriptaddress]{revtex4}
\usepackage{bbm}
\usepackage{amssymb}
\usepackage{amstext}
\usepackage{amscd}
\usepackage{amsthm}
\usepackage{amsopn}
\usepackage{amsmath}
\usepackage{amsbsy}
\usepackage{indentfirst}
\usepackage{graphicx}
\usepackage{verbatim}
\usepackage{bm}
\usepackage[bookmarks=false]{hyperref}
\newcommand{\dd}{\mathrm{d}}
\begin{document}
\author{Sung-Po Chao}
\affiliation{Center for Materials Theory, Department of Physics and
Astronomy, Rutgers University, Piscataway, NJ 08854}
\author{Guillaume Palacios}
\affiliation{Center for Materials Theory, Department of Physics and
Astronomy, Rutgers University, Piscataway, NJ 08854}
\affiliation{Instituut voor Theoretische Fysica, Universiteit van Amsterdam, Valckenierstraat 65, 1018 XE Amsterdam, The Netherlands}

 \title{Non-equilibrium Transport in the Anderson model of a biased Quantum Dot:\\
 Scattering Bethe Ansatz Phenomenology}

\begin{abstract}
We derive the transport properties of a quantum dot subject to
a source-drain bias voltage at zero temperature and magnetic field. Using
the Scattering Bethe Anstaz, a generalization of the traditional
Thermodynamic Bethe Ansatz to open systems out of
equilibrium, we derive results for the quantum dot occupation in and out of
equilibrium and, by introducing \emph{phenomenological} spin- and charge-fluctuation distribution functions
in the computation of the current, obtain the differential conductance
for large $\frac{U}{\Gamma}$.
The Hamiltonian to describe the quantum dot system is the
Anderson impurity Hamiltonian
and the current and dot occupation as a function of voltage are obtained numerically.
We also vary the gate voltage and study
the transition from the mixed valence to the Kondo regime in the
presence of a non-equilibrium current. We conclude with the difficulty we encounter
in this model and possible way to solve them without resorting to a \emph{phenomenological} method.
\end{abstract}
\pacs{72.63.Kv, 72.15.Qm, 72.10.Fk}
\maketitle
\section{Introduction}
The past few years have witnessed a spectacular progress in the
fabrication and exploration of nano-structures giving
experimentalists unprecedented control over the microscopic
parameters governing the physics of these systems. Nano-structures,
beyond their practical applications, display an array of emergent
phenomena stemming from their reduced dimensionality which enhances
quantum fluctuations and strong correlations. Often, experiments are
carried out under non-equilibrium conditions, with currents passing
through the structures. The measurements are performed over a wide
range of parameters, such as temperature and applied bias, allowing
experimental exploration of the interplay between non-equilibrium
dynamics and strong correlation physics \cite{Goldhaber-Gordon et
al, Cronenwett et, Schmid et al, SM, WG, Park}. A canonical example is the
non-equilibrium Kondo effect observed in a quantum dot attached to
two leads held at different chemical potentials $\mu_i$. The voltage
difference $V=\mu_1-\mu_2$ induces a non-equilibrium current $I(V)$ through
the dot, interfering with and eventually destroying the Kondo effect
as the voltage is increased.

 The standard theoretical description of the transport trough a quantum dot is
the two-lead Anderson impurity model under a bias voltage. The 1- or 2-lead
Hamiltonian at zero bias is exactly solvable via Bethe Ansatz \cite{PB,Kawakami}. Using this exact solution as well
as NRG calculations for example, the thermodynamics of the model have been studied
in great detail. But the non-equilibrium situation, namely when the two leads experience each a different chemical potential, is much more difficult. This is due to the subtle interplay between the 
non-equilibrium aspect of the problem on one hand and the presence of strong interactions on 
the other hand. Technically speaking it is very non trivial task to find a basis of states that
diagonalize simultaneously the voltage term and the interaction term in the Hamiltonian. 
Nevertheless, a lot of efforts have been put forward to study this model but so far, only approximate ways of dealing with the voltage and/or the interactions have been developed\cite{Ng,Raikh,Meir,Glazman,Rosch,Kristian,Luis,Schoeller,Eckel,Oguri,Oguri2,Zurab,Dagotto,Rapha,Daichi,Millis}.

 In this paper we develop a \emph{phenomenological} approach to the problem, based on 
 the Scattering Bethe Ansatz (SBA), recently developed by P. Mehta and N. Andrei (MA)
\cite{Mehta}, a
non-perturbative implementation of the Keldysh formalism to
 construct the current-carrying, \emph{open-system} scattering eigenstates for the
two-lead nonequilibrium Anderson impurity mode. The basic idea of the SBA is to construct scattering eigenstates of the full Hamiltonian defined directly on the infinite line and match the incoming states by two Fermi seas describing the initial state of the leads.
The non-equilibrium steady state transport properties of
 the system are then expressed
 as expectation values of the current or dot occupation operators in these eigenstates. This program has been implemented for the Interacting Resonance Level Model (IRLM), a spinless interacting model, described in Ref.~\onlinecite{Mehta} where the
  zero temperature results for current and dot occupation $\langle \hat{n}_d \rangle$ for all bias voltages were presented. Another exact solution of this model at the so-called self-dual point \cite{Schiller-Andrei} by E. Boulat, H. Saleur and P. Schmitteckert in Refs.~\onlinecite{Boulat1,Boulat2} uses conformal
  field theory techniques and compares successfully with t-DMRG results. 
  
  The main motivation of the present paper is to test the very interesting ideas behind
  the SBA framework on a physically more relevant model such as the Anderson impurity model
  and to focus on the phenomenology that can be extracted from it. Carrying out the program for the non-equilibrium Anderson model we find difficulties in the direct application of the SBA approach due to the fact that the ground state in the Bethe basis consists of bound pairs of quasi-particles, leading to problems in the computation of the scattering phase shifts for the quasi-particles with complex momenta. This problem is not present in the IRLM when the Bethe momenta are below the impurity level and no bound states can be formed. We circumvent this difficulty by means of the following argument: The transport property computed in the IRLM is related to the single particle phase shift across the impurity in the Bethe basis. Based on the same idea we develop a phenomenological approach to describe the transport property in the Anderson impurity model. We identify two types of possible
phase shifts across impurity, which we refer to as  "spin-fluctuation" and "charge-fluctuation"  types to label two phenomenological phase shifts akin to the fundamental excitations described in the traditional Bethe Ansatz in this model. The phenomenological Ansatz is checked against exact results on the dot occupation in equilibrium and the Friedel sum rule~\cite{Ambegaokar-Langer,Langreth}, in the linear response regime. Subsequently, we discuss our results for the out of equilibrium current, conductance and dot occupation. The scaling relations for the conductance, predicted from the
 Fermi liquid picture of the problem at strong and weak coupling, are also discussed.

The paper is organized as follows. We start with a formal construction of
scattering eigenstates in the two-lead Anderson impurity model. Then we discuss how we impose
boundary conditions, which serve as initial condition in the time dependent picture, on the electrons within the leads. Next we shall discuss our results for
the dot occupation in equilibrium and the conductance in the linear response regime. Based on the checks in equilibrium we then extend our computation to the out of equilibrium regime. The difficulty we encounter for complex momenta and the way we handle it will also be addressed there. Comparison with another attempt of exact solution for this model by R. M. Konik et al.~\cite{konik2,konik} with the idea of dressed excitations above Fermi energy in the Bethe Ansatz picture, first considered for the exact conductance of point contact device in the FQHE regime\cite{Saleur1,Saleur2}, will be discussed. We will also comment on the validity and implication of our numerical results, among them the charge susceptibility, in the out-of-equilibrium regime. Qualitative agreement between our theory and experimental result is then presented. The limit of $U\rightarrow \infty$ is also summarized in the last section based on the same phenomenological approach.
Finally, we summarize our results and conclude with some issues on the SBA approach to this model, and state how they could be overcome.


\section{The Scattering Bethe Ansatz approach}
	\subsection{Scattering state construction}
	In this section we apply the SBA approach to construct the
	scattering states of the full Hamiltonian.
	The (unfolded) 2-lead Anderson impurity Hamiltonian reads,
	\begin{multline}
	\hat{H}=\sum_{i=1,2}\int \dd x\,
	\psi^{\dagger}_{i\sigma}(x)(-i\partial_x)\psi_{i\sigma}(x)+
	\epsilon_d d^{\dagger}_\sigma
	d_\sigma\\
	+t_i(\psi^{\dagger}_{i\sigma}(0)d_\sigma+d^{\dagger}_
	\sigma\psi_{i\sigma}(0))+U\,d^\dagger_{\uparrow} d_\uparrow
	d^\dagger_\downarrow d_\downarrow \label{UAn}
	\end{multline}
	where summation over the spin indices $\sigma$ is implied. The fields
	$\psi_{i\sigma}(x)$ describe chiral, right-moving electrons from
	lead $i$, $U$ is the on-site Coulomb repulsion between electrons on
	the dot, $t_i$ is the coupling between the dot and the lead $i$, and
	$\epsilon_d$ is the gate voltage. We have set the Fermi velocity $v_F=1$.

	The model's equilibrium properties have been studied in great
	detail via the traditional Thermodynamic Bethe Ansatz (TBA)\cite{PB, Kawakami}.
	The SBA exploits in a new way the integrability of the Anderson
	Model to construct current-carrying scattering eigenstates on the
	open line. There are two main requirements: One is the construction
	of scattering eigenstates with the number of electrons in each lead conserved prior to scattering off the impurity.
	Another is the asymptotic boundary
	condition: that the wave function of the incoming electrons, i.e. in
	the region ($ x \ll 0$), tend to that of two free Fermi seas far from
	the impurity \cite{Mehta}. All information about the external bias
	applied to the system is encoded in the boundary condition by
	appropriately choosing the chemical potential of the incoming Fermi
	seas. As in all Bethe-Ansatz constructions, the full multi-particle
	 wavefunction is constructed from  single particle eigenstates  (now on the infinite open line)
	 and the appropriate
	two-particle S-matrices. We first rewrite Eq.~(\ref{UAn}) in the even-odd basis as
	\begin{eqnarray*}
	\hat{H}&=&\hat{H}_e+\hat{H}_o\\
	\hat{H}_e&=&\sum_{\sigma}\int \dd x\, \psi^{\dagger}_{e\sigma}(x)(-i\partial_x)\psi_{e\sigma}(x)+\epsilon_d d^{\dagger}_	\sigma d_\sigma \\
	 & &+\ t(\psi^{\dagger}_{e\sigma}(0)d_\sigma
	+d^{\dagger}_\sigma\psi_{e\sigma}(0))+Ud^	\dagger_{\uparrow}
	d_\uparrow d^\dagger_\downarrow d_\downarrow\\
	\hat{H}_o&=&\sum_{\sigma}\int \dd x\, \psi^{\dagger}_{o\sigma}(x)(-i\partial_x)\psi_{o\sigma}(x)
	\end{eqnarray*}
	With
	\begin{eqnarray*}
	\psi_{e\sigma}(x)&=&\frac{t_1\psi_{1\sigma}(x)+t_2\psi_{2\sigma}(x)}{\sqrt{t_1^2+t_2^2}}\\
	\psi_{o\sigma}(x)&=&\frac{t_2\psi_{1\sigma}(x)-t_1\psi_{2\sigma}(x)}{\sqrt{t_1^2+t_2^2}}
	\end{eqnarray*}
	and  $t=\sqrt{t_1^2+t_2^2}$.
 	In what follows we consider the case $t_1=t_2=\frac{t}{\sqrt{2}}$ for simplicity.
	The single particle solution for even and odd basis is:
	$|e,p\sigma\rangle=\int \dd x\, (e^{i px} g_p(x)\psi_{e\sigma}^\dagger(x)+e_p \delta(x)d_\sigma^\dagger)|0\rangle$
	and $|o,p\sigma\rangle=\int \dd x\, e^{i px} h_p(x)\psi_{o\sigma}^\dagger(x)|0\rangle$, with
	 $|0\rangle$ the vacuum state and $g_p(x)$, $h_p(x)$, $e_p$ independent of spin
	and given by
	\begin{eqnarray}
	&&g_p(x)=\theta(-x)+e^{i\delta_p}\theta(x)+s_{ep}\,\theta(x)\theta(-x)\ ,\nonumber\\
	\label{single}
	&&h_p(x)=\theta(-x)+\theta(x)+s_{op}\,\theta(x)\theta(-x)\ ,\\
	&&e_p=\frac{t(1+e^{i\delta_p}+s_{ep}/2)}{2(p-\epsilon_d)}\ .\nonumber
	\end{eqnarray}
	Here
	$\delta_p\equiv2\tan^{-1}(\frac{\Gamma}{\epsilon_d-p})$
  	is the usual single particle scattering phase shift of the electrons
	off the impurity obtained when setting $s_{ep}=0$. $\Gamma\equiv\frac{t^2}{2}$ is the width of the resonance level.
	We adopted a symmetric regularization scheme $\theta(\pm x)\delta(x)=\frac{1}{2}\delta(x)$
	and imposed $|p| \le D$, $D$ being the bandwidth cut-off \cite{AFL}. The $s(x)=\theta(x)\theta(-x)$ term is a local constant ($\partial_x s(x)=0$) in this scheme and it is included in the odd channel function to allow the same two particle S-matrices, Eq.(\ref{SmatU}), in all channels\cite{foot1,foot3}. The
	 $\theta(x)\theta(-x)$ term in the even channel wave function is introduced in order to modify
	 the original (when $s_{ep}=0$) single particle phase shift across the impurity.
	The choice of $s_{op}$ and $s_{ep}$ will be addressed later.
	In the lead basis, $|i,p \sigma\rangle$, the single-particle scattering eigenstates with the
	incoming particle incident from lead $i$, can be restored by taking a proper linear combination of even-odd states.
	For example, $|1,p\sigma\rangle=\frac{1}{\sqrt{2}}(|e,p\sigma\rangle+|o,p\sigma\rangle)=\int \dd 
	x\, e^{ip x}\alpha^{\dagger}_{1,p\sigma}(x)|0\rangle$ is written as
	\begin{multline}\label{single}
 	|1,p\sigma\rangle=\int \dd x \, e^{ipx}\Big\{
	[\theta(-x) + \frac{1}{2}(e^{i\delta_p}+1)\theta(x)]\psi_{1
	\sigma}^\dagger (x)\\
	+ \frac{1}{2}(e^{i\delta_p}-1)\theta(x)\psi_{2 \sigma}^\dagger(x) +
	e_p d_{\sigma}^\dagger \delta(x)+s^{\dagger}_{1p\sigma}(x)\Big\} |0\rangle
	\end{multline}
	with $|2,p \sigma\rangle=\frac{1}{\sqrt{2}}(|e,p\sigma\rangle-|o,p
	\sigma\rangle)=\int \dd x\, e^{ip x}\alpha^{\dagger}_{2,p\sigma}(x)|0\rangle$ and $s_{ip\sigma}^\dagger (x)$ related to the $\theta(x)\theta(-x)$ terms by 
	\begin{multline*}
	s^\dagger_{1p\sigma}(x)= \left(\frac{s_{ep}+s_{op}}{\sqrt{2}}\psi_{1\sigma}^\dagger(x)+\frac{s_{ep}-s_{op}}{\sqrt{2}}\psi_{2\sigma}^\dagger(x)\right)\times\\
	\times\theta(x)\theta(-x)
	\end{multline*}
	and 
	\begin{multline*}
	s^\dagger_{2p\sigma}(x)=\left(\frac{s_{ep}-s_{op}}{\sqrt{2}}\psi_{1\sigma}^\dagger(x)+\frac{s_{ep}+s_{op}}{\sqrt{2}}\psi_{2\sigma}^\dagger(x)\right)\times\\
	\times\theta(x)\theta(-x).
	\end{multline*}
	
	These states have a single incoming particle ($x<0$) from lead $i$, that is reflected back into lead $i$
	with amplitude, $R_p=(e^{i\delta_p}+1)/2$ and transmitted to the
	opposite lead with amplitude $T_p=(e^{i\delta_p}-1)/2$. Similar
	single particle states are discussed in Ref. \onlinecite{Mehta}.

	The multi-particle Bethe-Ansatz wave-function is constructed by means
	of the two-particle S-matrix, $\bm{\mathrm{S}}(p,k)$, describing the scattering of
	two electrons with momenta $p$ and $k$. By choosing $s_{op}=-4$ in Eq.~(\ref{single})  (the choice of $s_{ep}$ will be discussed in section B and does not affect the result here)
	in the single particle states we can construct the same two-particles S-matrix for all combinations in even-odd basis (see Appendix.~\ref{A-1}). The two-particles solution for both particles coming from lead $1$ in spin
	singlet state takes the following form	
\begin{multline*}
|1k,\uparrow; 1p,\downarrow\rangle\\
=\int \dd x_1\dd x_2 \mathcal{A}\{e^{i(kx_1+px_2)}Z_{kp}(x_1-x_2)\alpha^\dagger_{1k,\uparrow}(x_1)\alpha^\dagger_{1p,\downarrow}(x_2)\}|0\rangle\\
=\Big\{\int \dd x_1 \dd x_2 A\{g(x_1,x_2)\psi^\dagger_{e\uparrow}(x_1)\psi^\dagger_{e\downarrow}(x_2)\\
+ h(x_1,x_2)\psi^\dagger_{o\uparrow}(x_1)\psi^\dagger_{o\downarrow}(x_2)
+ j(x_1,x_2)(\psi^\dagger_{e\uparrow}(x_1)\psi^\dagger_{o\downarrow}(x_2)\\
-\psi^\dagger_{e\downarrow}(x_1)\psi^\dagger_{o\uparrow}(x_2))\}+\int \dd x A( e(x)(\psi^\dagger_{e\uparrow}(x)d^\dagger_\downarrow-\psi^\dagger_{e\downarrow}(x)d^\dagger_\uparrow)\\
+o(x)(\psi^\dagger_{o\uparrow}(x)d^\dagger_\downarrow-\psi^\dagger_{o\downarrow}(x)d^\dagger_\uparrow))+A m\, d^\dagger_\uparrow d^\dagger_\downarrow \Big\}|0\rangle
\end{multline*}
Here $\mathcal{A}$ is the antisymmetrizer. $A$ is an overall normalization factor and 
\begin{eqnarray*}
g(x_1,x_2)&=&Z_{kp}(x_1-x_2)g_k(x_1)g_p(x_2)\\&&+Z_{kp}(x_2-x_1)g_k(x_2)g_p(x_1)\\
j(x_1,x_2)&=&Z_{kp}(x_1-x_2)g_k(x_1)h_p(x_2)\\&&+Z_{kp}(x_2-x_1)h_k(x_2)g_p(x_1)\\
h(x_1,x_2)&=&Z_{kp}(x_1-x_2)h_k(x_1)h_p(x_2)\\&&+Z_{kp}(x_2-x_1)h_k(x_2)h_p(x_1)\\
e(x)&=&Z_{kp}(-x)g_p(x)e_k+Z_{kp}(x)g_k(x)e_p\\
o(x)&=&Z_{kp}(-x)h_p(x)e_k+Z^{eo}_{kp}(x)h_k(x)e_p\\
m&=&\tilde{Z}_{kp}(0)e_ke_p
\end{eqnarray*}
with $Z_{kp}(x)=e^{-i\phi_{kp}}\theta(-x)+e^{i\phi_{kp}}\theta(x)$ and $\tilde{Z}_{kp}(0)\equiv\frac{k+p-2\epsilon_d}{k+p-U-2\epsilon_d}Z_{kp}(0)$. Here $\tan(\phi_{kp})=\frac{-Ut^2}{(k-p)(p+k-U-2\epsilon_d)}$. The derivation and more general form of two particles case is written in Appendix.~\ref{A-1}. To include spin triplet case we denote $Z_{k_i,k_j}(x_i-x_j)\equiv Z_{k_i,k_j}(x_i-x_j)_{a_{i}a_{j}}^{a_{i}^{'}a_{j}^{'}}= \bm{\mathrm{I}}_{a_{i}a_{j}}^{a_{i}^{'}a_{j}^{'}}\theta(x_j-x_i)+\bm{\mathrm{S}}_{a_{i}a_{j}}^{a_{i}^{'}a_{j}^{'}}(k_i,k_j)\theta(x_i-x_j)$ where $a_i$ is the spin index before the scattering and $a_{i}^{'}$ the spin index after the scattering. $\bm{\mathrm{I}}_{a_{i}a_{j}}^{a_{i}^{'}a_{j}^{'}}$ is the identity matrix.
The S-matrices must satisfy
	the Yang-Baxter equations
\begin{eqnarray}\nonumber
&&\bm{\mathrm{S}}_{a_{1}a_{2}}^{a_{1}^{'}a_{2}^{'}}(k_1,k_2)\bm{\mathrm{S}}_{a_{1}a_{3}}^{a_{1}^{'}a_{3}^{'}}(k_1,k_3) \bm{\mathrm{S}}_{a_{2}a_{3}}^{a_{2}^{'}a_{3}^{'}}(k_2,k_3)\\\nonumber
&&=\bm{\mathrm{S}}_{a_{2}a_{3}}^{a_{2}^{'}a_{3}^{'}}(k_2,k_3)\bm{\mathrm{S}}_{a_{1}a_{3}}^{a_{1}^{'}a_{3}^{'}}(k_1,k_3)\bm{\mathrm{S}}_{a_{1}a_{2}}^{a_{1}^{'}a_{2}^{'}}(k_1,k_2)
\end{eqnarray}
for such a construction to be consistent. The two-particles S-matrix for this two-lead Anderson model is given by
	\begin{equation}
	\bm{\mathrm{S}}_{a_{i}a_{j}}^{a_{i}^{'}a_{j}^{'}} (k,p)=\frac{(B(k)-B(p))\bm {\mathrm{I}}_{a_{i}a_{j}}^{a_{i}^{'}a_{j}^{'}} +i2U\Gamma\, \bm{\mathrm{P}}_{a_{i}a_{j}}^{a_{i}^{'}a_{j}^{'}}}
	{B(k)-B(p)+i2U\Gamma}
	\label{SmatU}
	\end{equation}
	 with $B(k)= k(k-2\epsilon_d-U)$, $\bm{\mathrm{P}}=
	 \frac{1}{2}(\bm{\mathrm{I}}\cdot\bm{\mathrm{I}}+\vec{\sigma}\cdot\vec{\sigma})$
	 the spin exchange operator with $a_i$ and $a_j$ representing the incoming spin indices. Since the S-matrix is the same for all even-odd combinations the
	S-matrix does not depend on the lead index $i$, and the
	number of electrons in a lead, $N_i$, can change only at the
	impurity site.
	 This circumstance allows us to construct the
	fully-interacting eigenstates of our Hamiltonian characterized by
	the incoming quantum numbers, $N_1$ and $N_2$ the numbers of
	incident electrons from lead 1 and 2 respectively. These quantum
	numbers are subsequently determined by the chemical potentials
	$\mu_1$ and $\mu_2$.

	To complete the construction of the SBA current-carrying, scattering eigenstate, $|\Psi, \mu_i\rangle$, we must still
	choose the "Bethe-Ansatz momenta"  $\{p_l\}_{l=1}^{N_1+N_2}$ of the single particles states to ensure that the
	incoming particles look like two Fermi seas in the region $x<0$. This requirement translates into a set of
	 "free-field" SBA equations for the Bethe-Ansatz
	momenta-density of the particles from the two leads \cite{Mehta}.
	The argument is as follows: Away from the impurity
	$|i,p\sigma\rangle$ reduces to $ \psi^{\dagger}_{i\sigma}(x)$ with the
	inter-particle S-matrix Eq.~(\ref{SmatU}) present. Thus
	the scattering eigenstates describing non-interacting electrons are
	in the Bethe basis
	 rather than in the Fock basis of plane waves. The existence of
	 many basis for the free electron is due to their linear spectrum
	 which leads to degeneracy of the energy eigenvalues. The wave
	 function
	 $e^{ip_1x_1+ip_2x_2}[\theta(x_1-x_2) \,+\, \bm{\mathrm{S}} \theta(x_2-x_1)]A$  is an eigenstate of the
	 \emph{free} Hamiltonian for any choice of $\bm{\mathrm{S}}$ with, in particular, $\bm{\mathrm{S}}=\mathbbm{1}$
	 defining the Fock basis and $\bm{\mathrm{S}}$ given in Eq.~(\ref{SmatU}) defining the Bethe basis.
	 The Bethe basis is the correct "zero order" choice of a basis in the degenerate energy space
	 required in order to turn on the interactions. We proceed to describe the leads (two free Fermi seas)
	 in this basis.

	 We consider the system at zero temperature and
	zero magnetic field in this paper. To describe the two Fermi seas on the leads translates to a set of
	 Bethe Ansatz
	equations whose solution in this case consists of complex conjugate pairs: $p^{\pm}(\lambda)=x(\lambda)\pm
	iy(\lambda)$ in the $\lambda$-parametrization
	\cite{Kawakami, PB, foot2} with
	\begin{eqnarray}
	x(\lambda)&=&\tilde{\epsilon}_d
	-\sqrt{\frac{\lambda+\tilde{\epsilon}_d^2+\sqrt{(\lambda+\tilde{\epsilon}_d^2)^2+U^2\Gamma^2}}{2}}\label{defx1}\\\nonumber
	y(\lambda)&=&-\sqrt{\frac{-(\lambda+\tilde{\epsilon}_d^2)+\sqrt{(\lambda+\tilde{\epsilon}_d^2)^2
	+U^2\Gamma^2}}{2}}.
	\end{eqnarray}
	with $\tilde{\epsilon}_d = \epsilon_d +U/2$.
	 Each member of a
	pair can be either in lead $1$ or in lead $2$, since the S-matrix is unity in the lead space. There
	are, therefore, two possible configurations for these bounded pairs.
	One possible way of forming bounded pairs is described by four types of complex solutions whose
	densities we denote $\sigma_{ij}(\lambda)$ with $\{ij\}=\{11,12,21,22\}$ indicating the incoming
	electrons from lead $i$ and lead $j$. The other possibility, which is perhaps more intuitive in comparing
	with the free electron in the Fock basis, is to include only $\{ij\}=\{11,22\}$.
	These two types of states give the same results when evaluating the expectation value of the dot occupation in equilibrium.
	However when we turn on the bias voltage, the results obtained from a 4-bound states
	description show some charge
	fluctuations even way below the impurity level which is not expected from the non-interacting ($U\rightarrow 0$) theory (shown in Appendix~\ref{A}).
	Thus we shall disregard the 4-bound states solution on physical grounds and focus on the 2-bound states description in the following discussion.

 	To describe in the Bethe basis the two leads as two Fermi seas filled up to
	$\mu_1$ and $\mu_2$,
	respectively, these densities must satisfy the SBA equations,
	\begin{multline}
	\label{bulk}
	2\sigma_{i} (\lambda) = -\frac{1}{\pi} \frac{\mathrm{d}x(\lambda)}{\mathrm{d}\lambda}\theta(\lambda-B_i)\\
	-\sum_{j=1,2}\int_{B_{j}}^{\infty}\mathrm{d}\lambda' K(\lambda-\lambda')\sigma_{j}(\lambda')
	\end{multline}
	with $K(\lambda)=\frac{1}{\pi}\frac{2U\Gamma}{(2U\Gamma)^2+\lambda^2}$.
	
	The SBA equations are derived from
imposing boundary condition in the free leads (incoming state) region and the value of momenta is connected with
spin rapidity $\lambda$ by using the quantum inverse scattering method.
The Bethe Ansatz equations solved with periodic boundary conditions at the free lead region with total number of particles $N$ ($N=N_1+N_2$ as sum of particle number from lead 1 and 2) and the total spin projection $S$ ($S=S_1+S_2=N/2-M$ with $M=M_1+M_2$ as number of down spin particles from lead 1 and 2) are given by
\begin{eqnarray}
&&e^{ik^l_jL}=\prod_{\alpha=1}^M\frac{B(k^l_j)-\lambda_\alpha+iU\Gamma}{B(k^l_j)-\lambda_\alpha-iU\Gamma}\label{BAE1}\\\nonumber
&&\prod_{l=1,2}\prod_{j=1}^{N_l}\frac{B(k^l_j)-\lambda_\alpha-iU\Gamma}{B(k^l_j)-\lambda_\alpha+iU\Gamma}=\prod_{\beta\neq\alpha}^{M}\frac{\lambda_\alpha-\lambda_\beta+2iU\Gamma}{\lambda_\alpha-\lambda_\beta-2iU\Gamma}
\end{eqnarray}
with total energy $E=E_1+E_2$ and $E_l=\sum_{j}k_j^l$ indicating the energy of the electrons within the lead $l$ at zero temperature.

The spectrum of Eq.(\ref{BAE1}) for one lead case has been analyzed by N. Kawakami and A. Okiji \cite{Kawakami} where they found that the ground state at zero temperature is composed of real $\lambda_i$ and complex $k^l_j$ in the thermodynamic limit for $U>0$. The same situation also occurs in the special limit where $U\rightarrow\infty$ where P. Schlottmann~\cite{Schl} has done also in the one lead case. The proof for two leads ground state is similar to the one lead case and is shown explicitly for the finite temperature calculation for the infinite U case in Ref.~\onlinecite{foot2}.

As has been mentioned above in the zero temperature zero magnetic field ground state all $\lambda_i$ are real (and distinct) and $k^l_j$ form bound state for $j=1,..,2M$ with bound state momenta given by the poles or zeros
in the S-matrix defined in Eq.~(\ref{SmatU})
\begin{eqnarray}
B(k^{l\pm}(\lambda_j))=\lambda_j\pm iU\Gamma=B(x(\lambda_j)\pm i y(\lambda_j))+\gamma^{\pm}(\lambda_j)\label{kmom}
\end{eqnarray}
where $\gamma^{\pm}=O(\exp(-L))$ and $x(\lambda)$ and $y(\lambda)$ are shown in the Eq.~(\ref{defx1}).

\begin{figure}[h	]
		\includegraphics[width=1\columnwidth, clip]{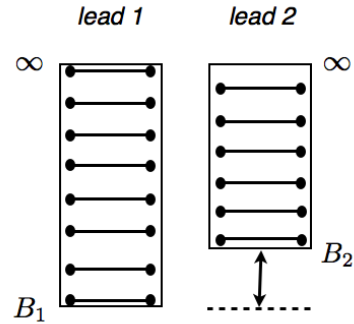}\\
		\caption{Sketch of the configuration of Bethe momenta corresponding to the ground state
		of $H_0$ with an additional bias voltage i.e. two Fermi seas at different chemical potential.
		 }
	  	\label{picSBA}
	\end{figure}

Note that the bound state can be formed from four possible configurations for $B_2<\lambda_\alpha<\infty$ which we denote bound state from lead $i$ and lead $j$ quasi momenta denoted as $\lambda^{ij}_\alpha$. The bound state between $B_1<\lambda_\alpha<B_2$ can only be formed by quasi momenta both coming from lead 1.
As already mentioned the four bound state distribution does not give physically sensible results for the charge susceptibility as shown in Appendix A and therefore we will limit our discussion to two types of bound state distribution here. Below we surpass the index of lead in $\lambda$ and put back the index dependence in the end for simplification.
Inserting Eq.~(\ref{kmom}) into Eq.~(\ref{BAE1}) we get
\begin{eqnarray}
&&e^{ik^+_\alpha L}=\prod_{\beta=1}^M\frac{\lambda_\alpha-\lambda_\beta+2iU\Gamma}{\lambda_\alpha-\lambda_\beta+\gamma_\alpha^+}\label{ab}\\
&&e^{ik^-_\alpha L}=\prod_{\beta=1}^M\frac{\lambda_\alpha-\lambda_\beta+\gamma_\alpha^-}{\lambda_\alpha-\lambda_\beta-2iU\Gamma}\label{ac}\\
&&\prod_{\beta=1}^M \frac{\lambda_\beta-\lambda_\alpha+\gamma_\beta^+}{\lambda_\beta-\lambda_\alpha+\gamma_\beta^-}=1\label{ad}
\end{eqnarray}
Thus for $L\rightarrow\infty$ from multiplication of Eq.~(\ref{ab}) and Eq.~(\ref{ac}) we have
\begin{eqnarray}
e^{2ix(\lambda_\alpha)L}=\prod_\beta\frac{\lambda_\alpha-\lambda_\beta+2iU\Gamma}{\lambda_\alpha-\lambda_\beta-2iU\Gamma}\label{BAE0}
\end{eqnarray}
Taking the logarithm of Eq.~(\ref{BAE0}) we have:
\begin{eqnarray}
2\pi J_\alpha=-2x(\lambda_\alpha)L- \sum_\beta \left(2\theta_2 \left(\frac{\lambda_\alpha-\lambda_\beta}{2U\Gamma}\right)+\pi \right)
\label{eqx}
\end{eqnarray}
with $\theta_n(x)\equiv \tan^{-1}(2x/n)$ and $\{J_\alpha\}$ a set of integer numbers. We can extend the definition of $J_\alpha$ to include integers or half integers and rewrite Eq.~(\ref{eqx}) as
\begin{eqnarray}
\frac{\pi}{L} J_\alpha=-x(\lambda_\alpha)- \frac{1}{L}\sum_\beta \theta_2 \left(\frac{\lambda_\alpha-\lambda_\beta}{2U\Gamma}\right)
\label{eqxnew}
\end{eqnarray}
Now let us put back the dependence in lead indices. Starting from Eq.~(\ref{eqxnew}) it can be shown that there is one-to-one correspondence between the $\lambda_\alpha$'s and the $J_\alpha$'s  and that all $\lambda_\alpha$'s have to be different. 
Thus the set of rapidities $\{\lambda^{ij}_\alpha\}$, characterizing an eigenstate of the Hamiltonian, is uniquely determined by one specific set of $\{J_\alpha\}$. For instance, the ground state of the Hamiltonian $H_0$ in 
the presence of a bias voltage is simply obtained by packing two "Fermi seas" of non-consecutive integers (Pauli principle in lead space) up to certain "Fermi points" (see Fig.~\ref{picSBA} ) corresponding to the 
$B_1$ and $B_2$ in the continuum limit. 
For notational simplification we relabel $\{ij\}=\{11,22\}$ as $\{l\}=\{1,2\}$.
Now defining $\sum_{ij}\sigma(\lambda^{ij}_\alpha)=\frac{1}{L}\frac{\dd J_\alpha}{\dd\lambda_\alpha}\equiv\sum_l\sigma^{(l)}(\lambda_\alpha)$ and using $\partial_x\theta_n(x)=\frac{2/n}{1+(2x/n)^2}$ we can write Eq.~(\ref{eqxnew}) in the continuum limit (by taking $L\rightarrow\infty$ and differentiate Eq.~(\ref{eqxnew}) with respect to $\lambda$). Doing so we shall distinguish two different domains:

For $B_2<\lambda<\infty$ the particles are fully packed and states are labeled by a different lead index 
$l$. In this domain, the SBA equations in the continuum limit takes the form
\begin{multline}
\sum_{l=1}^2\sigma^{(l)}(\lambda)=-\frac{1}{\pi}\frac{\dd x(\lambda)}{\dd\lambda}
-\int_{B_2}^{\infty} \dd\lambda'\, K(\lambda-\lambda')\sigma^{(2)}(\lambda')\\
-\int_{B_1}^{\infty} \dd\lambda'\, K(\lambda-\lambda')\sigma^{(1)}(\lambda')\ .
\label{finiU1}
\end{multline}

For $B_1<\lambda<B_2$ we can see from Fig.~\ref{picSBA} that the lead 2 states are unoccupied. 
We shall introduce a distribution of holes for the lead 2 that we will denote $\tilde{\sigma}^{(2)}(\lambda)$. The continuum SBA equations in this regime are given by
\begin{multline}
\sigma^{(1)}(\lambda)+\tilde{\sigma}^{(2)}(\lambda)=
-\frac{1}{\pi}\frac{\dd x(\lambda)}{\dd\lambda}-\int_{B_2}^{\infty} \dd\lambda'\,  
K(\lambda-\lambda')\sigma^{(2)}(\lambda')\\
-\int_{B_1}^{\infty} 
\dd\lambda'\, K(\lambda-\lambda')\sigma^{(1)}(\lambda') 
\label{finiU2}
\end{multline}
Since $\tilde{\sigma}^{(2)}(\lambda)$ obeys the same equation as $\sigma^{(2)}(\lambda)$ as may
 be seen from subtracting Eq.~(\ref{finiU2}) and Eq.~(\ref{finiU1}) we can combine Eq.~(\ref{finiU1}) and Eq.(\ref{finiU2}) together to get
\begin{multline}
2\sigma(\lambda)=-\frac{1}{\pi}\frac{\dd x(\lambda)}{\dd\lambda}
-2\int_{B_2}^{\infty} \dd\lambda'\, K(\lambda-\lambda')\sigma(\lambda')\\
-\int_{B_1}^{B_2} \dd\lambda'\, K(\lambda-\lambda')\sigma(\lambda')
\label{finiU3}
\end{multline}
with $B_1<\lambda<\infty$ for lead 1 and $B_2<\lambda<\infty$ for lead 2 Bethe momenta density distributions.
	Each density is defined on a domain extending from $B_{i}$ to the cutoff $D$ - to be sent to infinity.
	The $B_{i}$ play the role of chemical potentials for the Bethe-Ansatz momenta and are determined
	from the physical chemical potentials of the two leads, $\mu_i$, by minimizing the \emph{charge free energy},
	\begin{equation}
	F=\sum_i (E_i-\mu_i N_i)=2\sum_i \int_{B_{i}}^{\infty}\mathrm{d}\lambda\,
	(x(\lambda)-\mu_i)\sigma_{i}(\lambda)
	 \label{freeenergy}
	\end{equation}
	with $\sigma_{1}$ the lead $1$ particle density and
	$\sigma_{2}$ the lead $2$ particle density. Note that $\sigma_{1}$ and $\sigma_{2}$ obeys the same
	integral equation Eq.~(\ref{bulk}) with different boundary ($\sigma_{1}(\lambda)$ with $\lambda\subset(B_1,\infty)$ and $\sigma_{2}(\lambda)$ with $\lambda\subset(B_2,\infty)$).
	Solving the SBA equations subject to the minimization of the charge
	free energy fully determines the current-carrying eigenstate,
	$|\Psi, \mu_i\rangle$ and allows for calculation of physical
	quantities by evaluating expectation value of the corresponding operators.
	In the following we shall discuss our results from equilibrium cases to non-equilibrium ones,
	starting with the expression for various expectation value of physical quantities.
	\subsection{Expectation value of current and dot occupation}
	For $\mu_1=\mu_2$ all $B_{i}$ are equal to some equilibrium boundary $B$
	fixed by the choice of $\mu_i$. The dot occupation is given by the expectation value
	$ \sum_\sigma \langle\Psi, \mu_i| d^\dagger_\sigma d_\sigma|\Psi, \mu_i\rangle$. Taking the limit $L\rightarrow\infty$
($L$ being the size of the lead) one can express $n_d$ as an integral over the density of $\lambda$ and the corresponding matrix element $\nu(\lambda)\simeq\frac{\langle p^+(\lambda)p^-(\lambda)|\sum_{\sigma}d^\dagger_{\sigma}d_{\sigma}|p^+(\lambda)p^-(\lambda)\rangle}{\langle p^+(\lambda)p^-(\lambda)|p^+(\lambda)p^-(\lambda)\rangle}$ taken to order $\frac{1}{L}$. Here the state 
$|p^+(\lambda)p^-(\lambda)\rangle$ denotes a pair (or bound states) of quasi-particles with complex momenta given by Eq.~(\ref{defx1}). The reason why $n_d$ is governed solely by one-bound state matrix
element instead of a complicated many-particle object is because Bethe wave-functions are orthogonal to each other for different pairs of Bethe momenta under the condition that the size of the leads $L$ taken to infinity.

	Here we address the different choice of $s_{ep}$ (with $s_{op}=-4$ fixed to have the same S-matrix in all channels) which gives rise to
 different forms of $\nu(\lambda)$. We shall first discuss the "\emph{natural}" choice $s_{ep}=0$ 
 (i.e. absence of $\theta(-x)\theta(x)$ terms) and show it reproduces the \emph{exact} result for the dot occupation in equilibrium. While
 in checking the steady state condition, i.e. $d \langle n_d \rangle/dt=0$, for out-of-equilibrium 
 situation, the choice of $s_{ep}=0$ fails. To remedy this issue we propose $s_{ep}\neq 0$ 
 (i.e introducing counter-intuitive $\theta(-x)\theta(x)$ terms) schemes to circumvent this difficulty. We check this proposed \emph{phenomenological} scheme in equilibrium
 against the \emph{exact} dot occupation obtained in $s_{ep}=0$ case in the second part of the discussion as a benchmark for our approach. First let us discuss the result for $s_{ep}=0$:

    (1)~$s_{ep}=0$: We choose $s_{ep}=0$ as in the case of the 1-lead Anderson impurity model. Denote $\nu(\lambda)=\nu^{SBA}(\lambda)$ in this choice. The dot
occupation expectation value in equilibrium is given by
	\begin{multline}
	n_d = \frac{\langle\Psi,\mu_1=\mu_2|\sum_{\sigma}\hat{d}_\sigma^\dagger\hat{d}_\sigma|\Psi,\mu_1=\mu_2\rangle}{\langle\Psi,	 \mu_1=\mu_2|\Psi,\mu_1=\mu_2\rangle}\\
	=2\int_{B}^{\infty}\dd\lambda\, \sigma(\lambda) \nu^{SBA}(\lambda)		
	\label{dd}
	\end{multline}
	where the factor $2$ in front of the integral accounts for the spin degeneracy.
	The matrix element of the operator $d^\dagger_\sigma d_\sigma$ in the SBA state is given by
	\begin{widetext}
	\begin{equation*}
	\nu^{SBA}(\lambda)= \frac{2\Gamma}{\tilde{x}^2(\lambda)+\tilde{y}^2_+ (\lambda)}\\
	+\frac{16y(\lambda)\Gamma^2}{[\tilde{x}^2 (\lambda)+\tilde{y}^2_- (\lambda)]
	[\tilde{x}^2 (\lambda) +\tilde{y}^2_+ (\lambda)]}
	\left(\frac{\tilde{x}(\lambda)}{2\tilde{x}(\lambda)-U}\right)^2.
	\end{equation*}
	\end{widetext}
	where we introduced, for simplified notations, the functions
	$\tilde{x}(\lambda)=x(\lambda)-\epsilon_d$ and
	$\tilde{y}_\pm (\lambda)=y(\lambda)\pm\Gamma$.
	
		Eq.~(\ref{dd}) can be proved to be exact by comparing it with the traditional Bethe Ansatz
		(TBA)
  result. In the latter, $n_d$ is computed as the integral of the impurity density.
  This observation that the SBA and TBA results for $n_d$ agree in equilibrium shows the connection between the dot occupation and the dressed phase shift across the impurity. The dressed phase shift mentioned here is equivalent to the impurity density as can be seen in the Eq.(\ref{ddT}) in Appendix \ref{B}.
The proof of the equivalence between TBA and SBA in equilibrium is also given in Appendix \ref{B}.

  	To describe the out-of-equilibrium state we first check if the steady state condition
	$\frac{d\langle\hat{n_d}\rangle}{dt}=0$ (or equivalently, $\frac{d\langle\hat{N_1}+\hat{N_2}\rangle}{dt}=0$) is satisfied in
	this basis. As mentioned earlier these scattering states are formed by bounded quasi-particles
	with complex momenta
	and therefore the single particle phase across the impurity is not well defined in the sense
	that $|e^{i\delta_{p^{\pm}}}|\neq 1$. This problem begins to surface as we set out to evaluate transport
	expectation value and renders
	\begin{eqnarray}
	\label{nonsteady}
	&&\frac{d\langle\hat{n_d}\rangle}{dt}=\int_{B_{11}}^{B_{22}}d\lambda\sigma_b(\lambda)\Delta(\lambda)\neq 0
	\end{eqnarray}
	with $$\Delta(\lambda)=\frac{y^2(\lambda)\Gamma^2}{[\tilde{x}^2(\lambda)
	+\tilde{y}_-^2(\lambda)][\tilde{x}^2(\lambda)+\tilde{y}_+^2(\lambda)]}\ .$$
	Thus it appears that using this basis the steady state condition is not observed. This problem does not appear when the momenta are real as in the IRLM case\cite{Mehta}.
		 
	 (2)~$s_{ep}\neq 0$: To remedy this problem we redefine the single particle phase shifts across the impurity, in analogy to the results for the
    IRLM\cite{Mehta}, through the choice of nonzero $s_{ep}$ in Eq.(\ref{single}). With a suitable choice of $s_{ep}$ we may restore
	a well defined single particle phase $|e^{i\tilde{\delta}_{p^{\pm}}}|=1$ with $\tilde{\delta}_{p^{\pm}}$ denoting
	this new phase. The way we judge whether we make the correct choice for the new phases $\tilde{\delta}_{p^{\pm}}$
is to compare the dot occupation $n_d$ in equilibrium before and after the redefined phase.
The explicit form of $s_{ep}$ and phase $\tilde{\delta}_{p^{\pm}}$ will be motivated below but first we shall show that a single redefined
phase is not sufficient to satisfy the constraint of dot occupation comparison.

	Again the choice of new phases is constrained by the requirement that we shall obtain the same
	result for $\langle\sum_\sigma d^\dagger_\sigma d_\sigma\rangle$ as given by $\nu^{SBA}(\lambda)$
	in equilibrium. Based on this constraint it can be shown explicitly that a single well defined phase (in the sense of $|e^{i\tilde{\delta}_{p^{\pm}}}|=1$)
    is not sufficient to reproduce the equilibrium $\nu^{SBA}(\lambda)$ as following:
	The new dot amplitude $\tilde{e}_{p^+}$ and $\tilde{e}_{p^-}$ have to satisfy
	\begin{eqnarray*}
	|\tilde{e}_{p^+}|^2+|\tilde{e}_{p^-}|^2 &=& \frac{4\Gamma}{\tilde{x}^2(\lambda)+
	\tilde{y}^2_+(\lambda))}\ ,\\
	|\tilde{e}_{p^+}|^2|\tilde{e}_{p^-}|^2 &=&
	\frac{4\Gamma^2}{[\tilde{x}^2(\lambda)
	+\tilde{y}^2_+(\lambda)][\tilde{x}^2(\lambda)+\tilde{y}^2_-(\lambda)]}\ .
	\end{eqnarray*}
	As both $|\tilde{e}_{p^+}|^2$ and $|\tilde{e}_{p^-}|^2$ are positive we see that a single redefined phase cannot satisfy the above constraints
simultaneously. Therefore we have to choose at least two sets of redefined phases $\tilde{\delta}_{p^{\pm}}^i$ (with $i=s,h$ denoting spin-fluctuation or charge-fluctuation to be addressed later) and, along with them,
	some distribution functions $f^i$ to set the weight for these phases.

	To motivate the idea of searching the correct
	phase shifts we shall come back to the derivation of dot occupation in traditional Bethe Ansatz (TBA) picture.
	In TBA the total energy of the system is described by energy of the leads electrons and energy shifts from
	the impurity,
	\begin{equation}
	\label{energy}
	E=\sum_{j}p_j=\sum_j \left(\frac{2\pi n_j}{L}+\frac{1}{L}\delta_j\right)
	\end{equation}
 	Based on Feynman-Hellman theorem, which is applicable in equilibrium (closed) system,
	we have
	\begin{equation}
	\label{FH}
	\langle\hat{n}_d\rangle=\frac{\partial E}{\partial \epsilon_d}=\frac{1}{L}\sum_j\frac{\partial \delta_j}{\partial \epsilon_d}=\frac	 {1}{L}\sum_j\frac{\partial (\delta_{p_j^+}+\delta_{p_j^-})}{\partial \epsilon_d}
	\end{equation}
	The result for Eq.~(\ref{FH}) agrees with those obtained from Eq.~(\ref{comtba}) and can be
	viewed as a third approach to obtain the expectation value of the dot occupation. The key
	observation here is that this
	quantity is related to the \emph{bare} phase shift $\delta_{p^+}+\delta_{p^-}$ and therefore the
	redefined phases must be proportional to this quantity. Among them there are two likely
	candidates
	with redefined phase shift given by $\delta_{p^+}+\delta_{p^-}$, describing the tunneling of a
	bounded pair,
	and $\frac{\delta_{p^+}+\delta_{p^-}}{2}$, describing the tunneling of a single quasi-particle.
	In a sense this is the
	echo for the elementary excitations above the Fermi surface in the Bethe basis
	characterized by N. Kawakami
	and A. Okiji\cite{Kawakami2} as charge-fluctuation excitation, which describes bounded pair
	quasi-particles excitation, and
	spin-fluctuation excitation, which describes one quasi-particle excitation.
	Another similar picture is the spin-fluctuation and charge-fluctuation two fluids picture proposed by D. Lee et al\cite{Lee} albeit in a different context.
	We identify the phase defined by
	$$\tilde{\delta}_{p^-}=\tilde{\delta}_{p^+}=\frac{\delta_{p^+}+\delta_{p^-}}{2}\equiv\tilde{\delta}_p^s$$
	(with $s_{ep^{\pm}}\equiv s_{ep^{\pm}}^s=\frac{2}{\Gamma}(i(p^{\pm}-\epsilon_d)-\Gamma)
	(e^{i(\frac{\delta_{p^+}+\delta_{p^-}}{2})}-1)$) as spin-fluctuation phase shift and
	$$\tilde{\delta}_{p^-}=\tilde{\delta}_{p^+}=\delta_{p^+}+\delta_{p^-}\equiv\tilde{\delta}_p^h$$
	(with $s_{ep^{\pm}}\equiv s_{ep^{\pm}}^h=\frac{2}{\Gamma}(i(p^{\pm}-\epsilon_d)-\Gamma)
	(e^{i(\delta_{p^+}+\delta_{p^-})}-1)$) as charge-fluctuation phase shift.

    The out-of-equilibrium current is evaluated by the expectation value of current operator $\hat{I}$
	with $\langle\hat{I}\rangle$ defined by
	\begin{equation}
	\label{curdef}
	\langle\hat{I}\rangle=\frac{-\sqrt{2}iet}{\hbar}
	\langle\sum_\sigma((\psi_{1\sigma}^{\dagger}(0^{\pm})-\psi_{2\sigma}^{\dagger}(0^{\pm}))d_\sigma-h.c.)\rangle
	\end{equation}
	 in the state $|\Psi, \mu_i\rangle$.
	Notice that $\psi_{i\sigma}^{\dagger}(0^{\pm})\equiv\lim_{\epsilon\rightarrow0}(\psi_{i\sigma}^{\dagger}(-\epsilon)
	+\psi_{i\sigma}^{\dagger}(+\epsilon))/2$ is introduced in transport related quantity to be consistent with our
	regularization scheme which introduces another local discontinuity in odd channel at impurity site.

  From Eq.~(\ref{curdef}) and the expression for the phases $\tilde{\delta}_p^s$ and $\tilde{\delta}_p^h$ we have the expression for current as
	\begin{multline}
	I(\mu_1,\mu_2)=  \langle\Psi, \mu_1,\mu_2|\hat{I}|\Psi,
	\mu_1,\mu_2\rangle  \\
	=\frac{2e}{\hbar}\int_{B_{1}}^{B_{2}}\mathrm{d}\lambda\ \sigma_b(\lambda)
	(f_s(\lambda)J^s(\lambda)+f_h(\lambda)J^h(\lambda))
	\label{current}
	\end{multline}

	The corresponding spin-fluctuation and charge-fluctuation matrix element of the current operator based on the spirit of Landauer transport, denoted as $J^s(\lambda)$
	and $J^h(\lambda)$ with $J^{\alpha}(\lambda)=|T_p(\lambda)|^2=|\frac{e^{i\tilde{\delta}_p^{\alpha}}-1}{2}|^2$ ($\alpha=\{s,h\}$) depending on redefined phase shift $\tilde{\delta}_p^{\alpha}$ only, are given by
	\begin{eqnarray}
	\label{Js}
	J^s(\lambda)&=&1+\frac{\mathrm{sgn}(\tilde{x}(\lambda))(\tilde{x}^2(\lambda)
	+y^2(\lambda)-\Gamma^2)}{\sqrt{(\tilde{x}^2(\lambda)+y^2(\lambda)-\Gamma^2)^2
	+4\Gamma^2 \tilde{x}^2(\lambda)}}\\
	\label{Jh}
	J^h(\lambda)&=&\frac{2\Gamma^2 \tilde{x}^2(\lambda)}
	{(\tilde{x}^2(\lambda)+\Gamma^2)^2-2y^2(\lambda)(\Gamma^2-\tilde{x}^2(\lambda))
	+y^4(\lambda)} .\nonumber\\
	&&
	\end{eqnarray}
	Here $\mathrm{sgn}(x)=\frac{x}{|x|}$ is the sign function. It is introduced in order to pick up the correct branch
    when taking the square root in denominator of Eq.~(\ref{Js}).
	This way we ensure that $J^s(\lambda)$ has the proper limit when $U$ is sent to infinity (cf Section III).
	Other than the motivations mentioned above for identifying spin and charge fluctuation phase shifts the functional forms of $J^s(\lambda)$ and
$J^h(\lambda)$ as a function of bare energy $x(\lambda)$ can also be used to identify these two type of phase shifts (See Fig.~\ref{mco} in Section III for infinite $U$ Anderson model, the finite $U$ is similar).
	
	Next we shall choose the appropriate weight for each type of phase shift.
	So far we have not yet been able to deduce the form
	of these weight functions $f_s(\lambda)$ and $f_h(\lambda)$
	and we introduce them \emph{phenomenologically}.
 	Let us define \emph{phenomenological} spin-fluctuation and charge-fluctuation weight functions as
	\begin{equation}
	\label{fs}
	f_s(\varepsilon(\lambda))=\frac{D_s(\varepsilon(\lambda))}{D_s(\varepsilon(\lambda))+D_h(\varepsilon(\lambda))}
	\end{equation}
	and
	\begin{equation}
	\label{fh}
	f_h(\varepsilon(\lambda))=\frac{D_h(\varepsilon(\lambda))}{D_s(\varepsilon(\lambda))+ D_h(\varepsilon(\lambda))}\ .
	\end{equation}
	Here $D_s(\varepsilon(\lambda))$ is the spin-fluctuation density of state, $D_h(\varepsilon(\lambda))$ is the charge-fluctuation density of state as defined in
	Ref.~\onlinecite{Kawakami2}, and $\varepsilon(\lambda)$ is the corresponding dressed energy i.e. the energy required to produce these spin- and charge-fluctuation excitations above the Fermi level.
Here dressed energy refers to the sum of the bare energy of adding/removing one bound state, as in charge fluctuation, or single quasi particle, as in spin fluctuation, and the energy shift from other quasi particles due to this change. The equation that solves a single quasi-particle's dressed energy $\varepsilon(\lambda)$
reads\cite{footnote3}
	\begin{equation}
	\varepsilon(\lambda)=(x(\lambda)-\mu)-\int_B^{\infty}\mathrm{d}\lambda'\, K(\lambda-\lambda')
	\varepsilon(\lambda')\, .
	\label{denergy}
	\end{equation}
	We wish to compare at this point our approach to the one taken by Konik et al\cite{konik2,konik}. The authors' Landauer approach is based on an ensemble of renormalized excitations, the holons and spinons, and the conductance is expressed in terms of their phase shift crossing the impurity. However, the leads are built of bare electrons and thus one faces the difficult problem of how to construct a bare electron out of renormalized excitations in order to be able to impose the voltage boundary condition. The basic approximation adopted, {\it electron $\approx$ antiholon + spinon}, is valid only when the electron is close to the Fermi surface (see N. Andrei~\cite{Andrei 82}), and therefore the approach is trustworthy only for very small voltages.
	Nevertheless, the dressed excitations framework seems to give at least qualitatively good results when another energy scale (such as the temperature or an external field) is turned on~\cite{GKLS}.
	In contrast
we construct the eigenstates of the Hamiltonian directly in terms of the bare electron field
and can therefore impose the asymptotic boundary condition that the wave function tend to a product of two free Fermi seas composed of bare electrons.
	While we do not have a mathematically rigorous derivation of the weight functions we introduced, the validity of the scattering formalism is not restricted to any energy window other than energy cutoff.

	\subsection{Results for equilibrium and linear response}
	In the numerical computation, for the practical purpose, we assumed Kondo limit ($U=-2\epsilon_d$, $\frac{U}{\Gamma}\gg 1$) form of the spin-fluctuation and
	charge-fluctuation distributions, i.e.
	\begin{equation}
	D_s(\varepsilon(\lambda))\simeq \frac{1}{\pi}\frac{T_k}{\varepsilon^2(\lambda)+T_k^2}
	\end{equation}
	and
	\begin{equation}
	D_h(\varepsilon(\lambda))\simeq\frac{1}{\sqrt{2U\Gamma}}\frac{\Gamma^2}{(\varepsilon(\lambda)+\epsilon_d)^2+\Gamma^2}
	\end{equation}
	with $T_k$ being the Kondo scale derived in Ref.~\onlinecite{Kawakami2} as
	\begin{equation}
	T_k=\frac{\sqrt{2U\Gamma}}{\pi}e^{\pi\frac{\epsilon_d(\epsilon_d+U)
	+\Gamma^2}{2U\Gamma}}\ .
	\label{TK}
	\end{equation}
	As we use the Kondo limit in our expression for the spin-fluctuation and
	charge-fluctuation distributions, we expect our phenomenological approach works better for large $U/\Gamma$.
	We also take
	$\varepsilon(\lambda)\simeq x(B)-x(\lambda)$ for numerical convenience with $B$ denoting the Bethe momenta boundary given by $\mu_1=\mu_2=0$.
	The dot occupation $\langle\sum_\sigma d^\dagger_\sigma d_\sigma\rangle$ evaluated by these new phases is given by
	\begin{multline}
	\label{dot occ}
	\langle\sum_\sigma d^\dagger_\sigma d_\sigma\rangle=2\Bigg(\int_{B1}^{\infty}\dd \lambda\,\sigma_b(\lambda)( \nu^s(\lambda)f_s(\lambda)+\nu^h(\lambda)f_h(\lambda))\\
	+\int_{B_2}^{\infty}\dd \lambda\, \sigma_b(\lambda)( \nu^s(\lambda)f_s(\lambda)
	+\nu^h(\lambda)f_h	(\lambda))\Bigg)
	\end{multline}	
	with $\nu^s(\lambda)$ and $\nu^h(\lambda)$ given as	
	\begin{widetext}
	\begin{multline}
	\nu^s(\lambda)=\frac{1}{\Gamma}\left[1-\frac{(\tilde{x}^2(\lambda)
	+y^2(\lambda)-\Gamma^2)}{\sqrt{(\tilde{x}^2(\lambda)+y^2(\lambda)-\Gamma^2)^2
	+4\Gamma^2 \tilde{x}^2(\lambda)}}\right]\\
	\times\left[1+8y(\lambda)\frac{1}{\Gamma}\left(1-\frac{(\tilde{x}^2(\lambda)
	+y^2(\lambda)-\Gamma^2)}{\sqrt{(\tilde{x}^2(\lambda)+y^2(\lambda)-\Gamma^2)^2
	+4\Gamma^2 \tilde{x}^2(\lambda)}}\right)\left(\frac{\tilde{x}(\lambda)}
	{2\tilde{x}(\lambda)-U}\right)^2 \right]
	\end{multline}
	\begin{multline}
	 \nu^h(\lambda)=\left[\frac{2\Gamma\tilde{x}^2(\lambda)}{(\tilde{x}^2(\lambda)+\Gamma^2)^2
	 -2y^2(\lambda)(\Gamma^2-\tilde{x}^2(\lambda))+y^4(\lambda)}\right]\\
	 \times\left[1+\frac{36 y(\lambda)\Gamma \tilde{x}^2(\lambda)}{(\tilde{x}^2(\lambda)+
	 \Gamma^2)^2-2y^2(\lambda)(\Gamma^2-\tilde{x}^2(\lambda))+y^4(\lambda)}
	 \left(\frac{\tilde{x}(\lambda)}{2\tilde{x}(\lambda)-U}\right)^2\right]
	\end{multline}
	\end{widetext}
respectively. We may check whether this choice of \emph{phenomenological distribution functions} satisfy the condition in equilibrium that
	\begin{multline}
	\label{condition}
	\langle\sum_\sigma d^\dagger_\sigma d_\sigma\rangle=4\int_{B}^{\infty}\dd\lambda\, \sigma_b(\lambda) \nu^{SBA}(\lambda)\\
	=4\left(\int_{B}^{\infty}\dd\lambda\, \sigma_b(\lambda)( \nu^s(\lambda)f_s(\lambda)
	+\nu^h(\lambda)f_h	(\lambda))\right)\ .
	\end{multline}
	 We can see from the Top of Fig.~\ref{eqdgplot} that the comparison between the phenomenological and the exact result for the dot occupation in equilibrium
     is good deep into the Kondo regime ($\epsilon_d\simeq-\frac{U}{2}$)
	 and far away from it ($\epsilon_d\gg 0$) but is worse when we are in mixed valence region ($\epsilon_d\simeq 0$).
	 This discrepancy, due in part to the approximations we made for $D_s(\varepsilon)$ and $D_h(\varepsilon)$,
	 may go away if we took more realistic form of $D_s(\varepsilon(\lambda))$ and
	 $D_h(\varepsilon(\lambda))$ also in mixed valence regime
	 as suggested in Fig.~\ref{eqdgplot}. However the numerical procedure is much more complicated there. We confine ourself to this simpler limit in
our phenomenological approach.

	Another check on our result in equilibrium is to find the linear response conductance through our
	formulation and compare with the exact linear result given by the Friedel sum rule\cite{Ambegaokar-Langer,Langreth}.
	The Friedel sum rule, which relates the equilibrium dot occupation to the phase
	shift experienced by electrons crossing the dot, is related to zero voltage conductance by
	$\frac{dI}{dV}|_{V=0}= 2 \sin^2(\pi \langle \hat{n}_d \rangle/2)$.
	The zero bias conductance in our construction can be analyzed easily\cite{footnote4} by noting that at low-voltage
	$eV=\mu_1-\mu_2\simeq\frac{2\pi}{L}(N_1-N_2)=4\pi\int_{B_{1}}^{B_{2}}\sigma_b(\lambda)d\lambda$.
	 By taking $B_{2}\simeq B_{1}=B$ in the expression for the current across the impurity Eq.~(\ref{current})
	we get the zero bias conductance expressed as
	\begin{equation}
	\frac{dI}{dV}\Big |_{V=0}=\frac{e^2}{h}\left[f_s(B)J^s(B)+f_h(B)J^h(B)\right]
	\label{zbc}
	\end{equation}
	Here $B=B(\mu,\epsilon_d,\Gamma,U)$ is determined by $\mu_1=\mu_2=0$.
	The comparison between Friedel sum rule (FSR) result and the conductance given by
	Eq.~(\ref{zbc})
	(denoted as (pSBA)) is shown at the Bottom of Fig.~\ref{eqdgplot}. It displays the consequence
	of the equilibrium Kondo effect in the quantum dot set up: due to the formation of the Kondo peak attached
	to the Fermi level the Coulomb blockade is lifted and a unitary conductance is reached for a range of gate
	voltages $\epsilon_d$ around $-U/2$. Again we see that the comparison is good for large $U/\Gamma$ but
	poorer in mixed valence regime for smaller $U/\Gamma$, which is consistent with the observation we made when
	evaluating $\langle \hat{n}_d\rangle$ as shown in top figure of Fig.~\ref{eqdgplot}. Having checked our results in equilibrium we shall go on
	to compute the current and the dot occupation in the out-of-equilibrium regime.
	\subsection{Results Out-Of-Equilibrium}
	\begin{figure}[t]
	\includegraphics[width=1\columnwidth, clip]{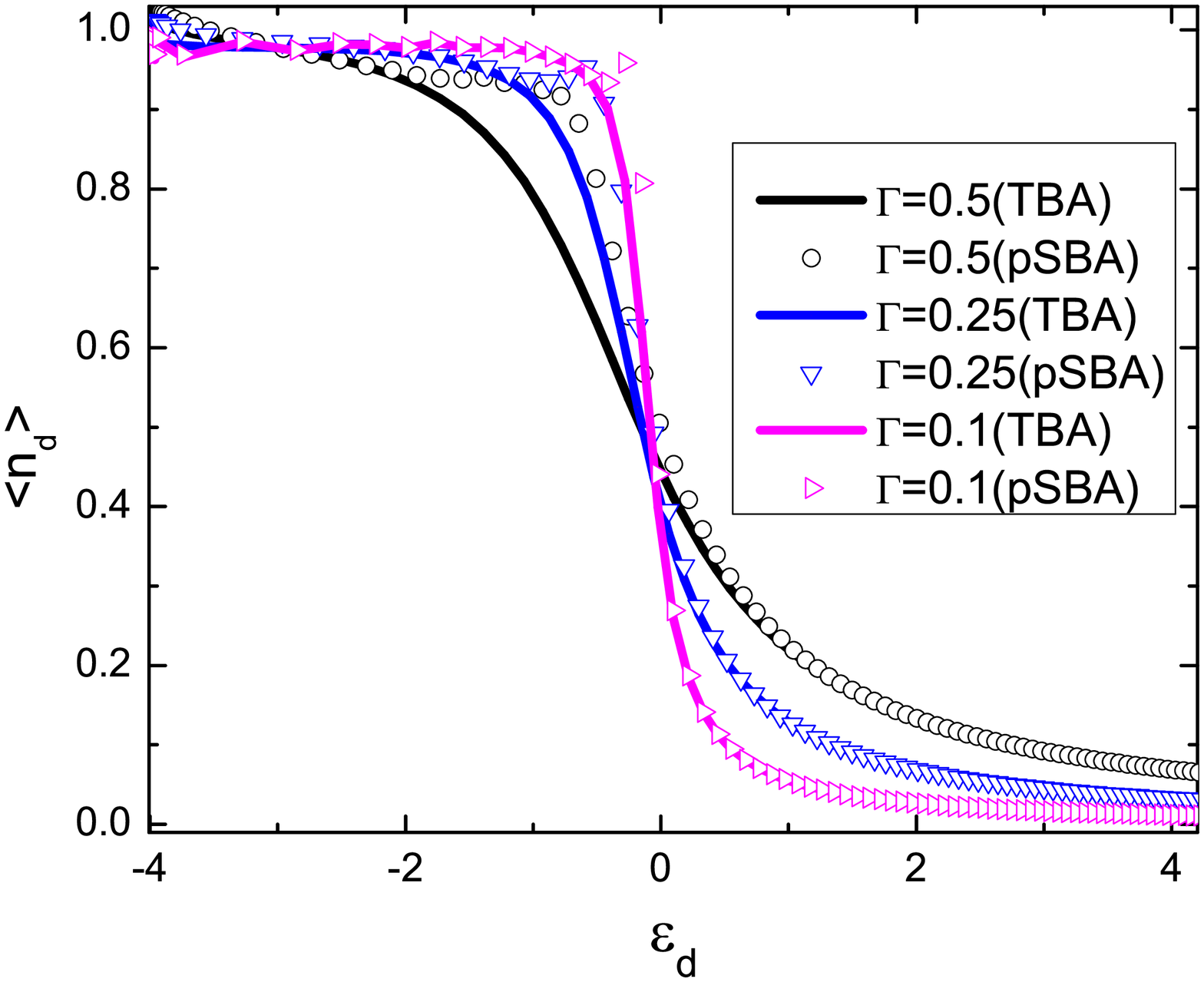}\\
	\includegraphics[width=1\columnwidth, clip]{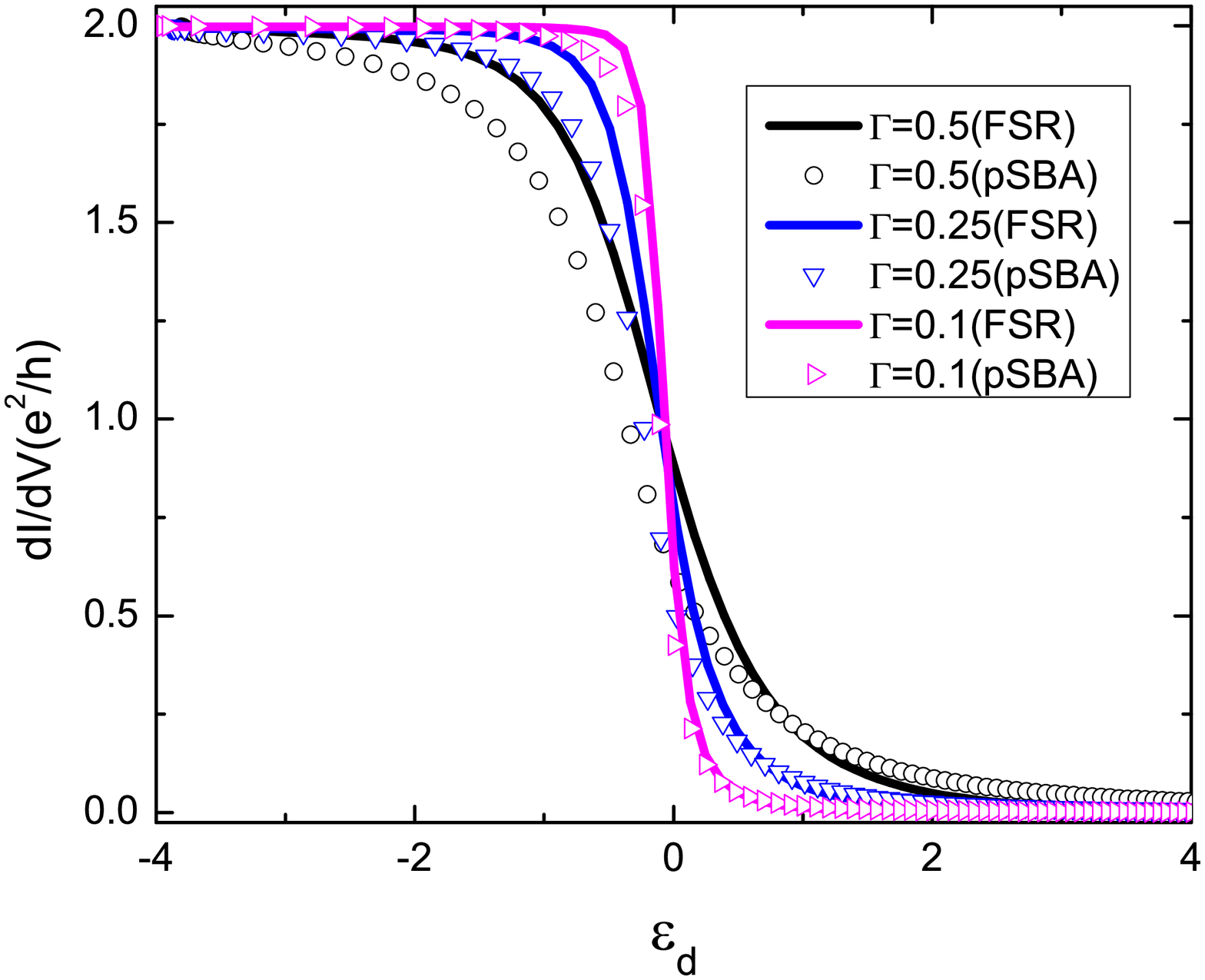}
	\caption{Top: $\langle\hat{n_d}\rangle$ as a function of $\epsilon_d$ from the exact result (dotted
	line) and from Eq.~(\ref{condition}) (solid line). Bottom: The
	 differential conductance in the linear-response regime, as a function of
	 $\epsilon_d$ from the phenomenological Scattering
	  Bethe Ansatz (pSBA) and exact linear response conductance from Friedel sum rule (FSR) for
	  $\Gamma=0.5$, $0.25$, $0.1$, and $U=8$.}
	  \label{eqdgplot}
	\end{figure}
\begin{figure}[h]
	\includegraphics[width=1\columnwidth, clip]{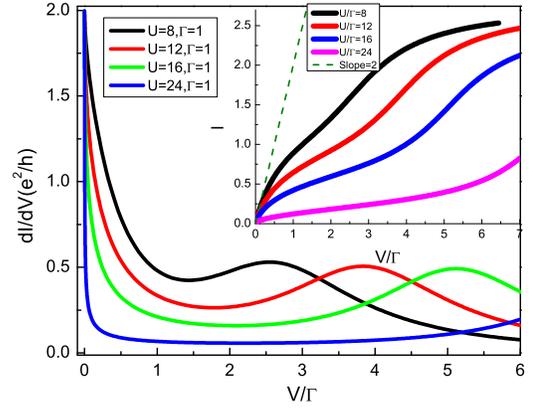}
\caption{$dI/dV$ vs $V/\Gamma$ for $\Gamma=1$, $\epsilon_d=-U/2$, and various $U$.
	  Inset: Steady state current vs voltage curves for  $\Gamma=1$, $\epsilon_d=-U/2$, and various $U$. Dashed line is a line with constant conductance $\frac{2e^2}{h}$ plotted for comparison.	
  	}\label{didvneqfig1}
\end{figure}

	Now let us begin to investigate the current and dot occupation change as we turn on the voltage.
	 We start with the discussion on current vs voltage for various regime.
	The current vs voltage is plotted in the inset of figure of Fig.~\ref{didvneqfig1} for different values
	 of $U$ and at the symmetric point $\epsilon_d = -U/2$. Note that we use an asymmetric
	 bias voltage when solving numerically the integral equations originating from Eq.~(\ref{bulk})
	 with constraint of minimizing the charge free energy Eq.~(\ref{freeenergy}):
	 Namely we fix $\mu_1\simeq 0$ (around $10^{-3}-10^{-5}$)
	 and lower $\mu_2$. Therefore, a direct confrontation between the results obtained from
	 real-time simulations of the Anderson model out-of-equilibrium \cite{Dagotto, Millis, Eckel} is difficult
	 but the main features of our calculation match the predicted results:
	 a linear behavior of the $I$-$V$ characteristics at low-voltage,
	 the slope being obtained from the FSR (2 in units of $e^2/h$ at the symmetric point), and a non-monotonic behavior at higher voltage, the
	 so-called non-linear regime. In particular, our calculations show clearly that the current will
	 decrease as $U/\Gamma$ is increased which is in agreement with other numerical approaches
	 (e.g. cf Fig. 2 of Ref.~\onlinecite{Dagotto} for a comparison).
	
\begin{figure}[h]
	\includegraphics[width=1\columnwidth, clip]{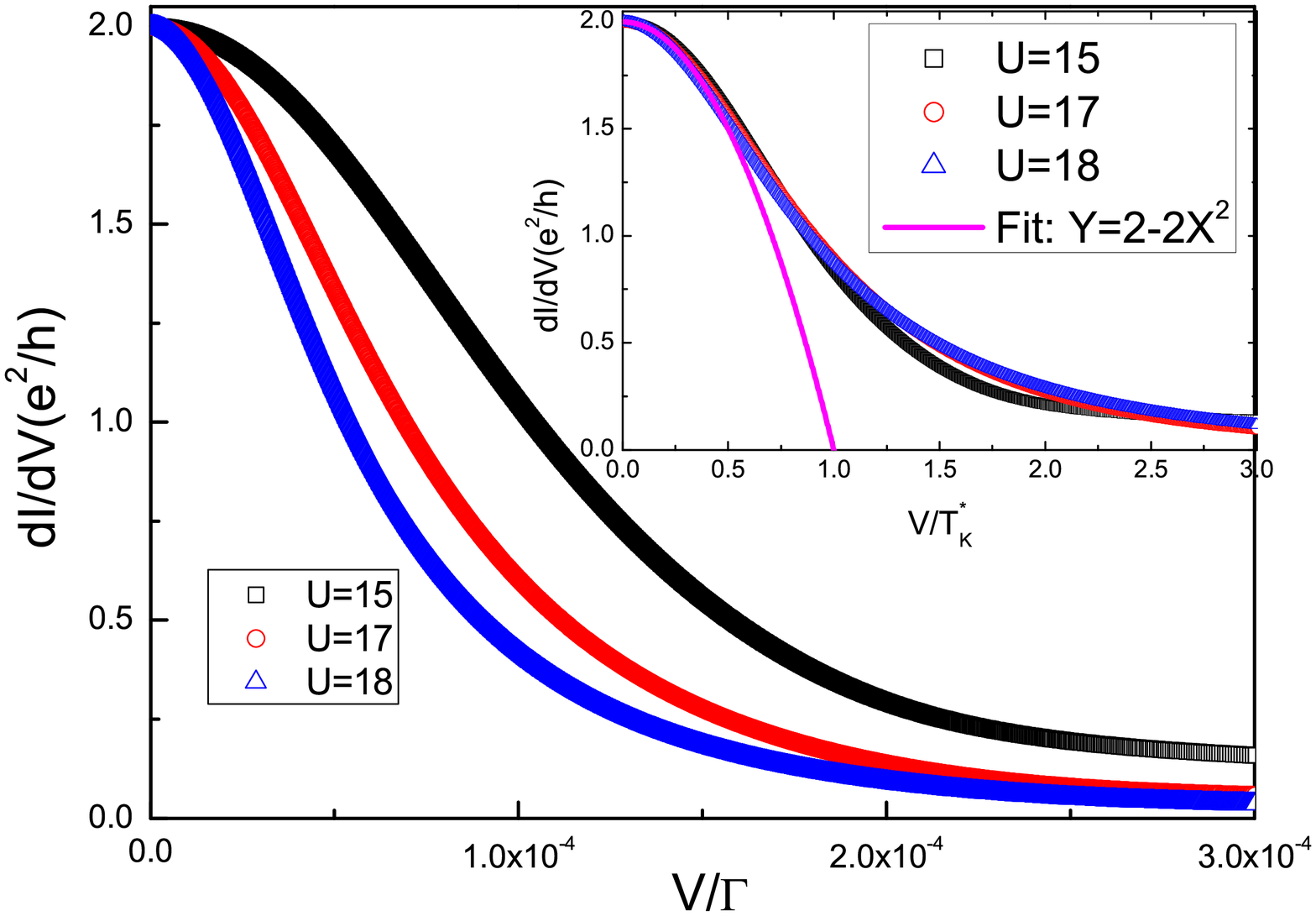}\\
	\includegraphics[width=1\columnwidth, clip]{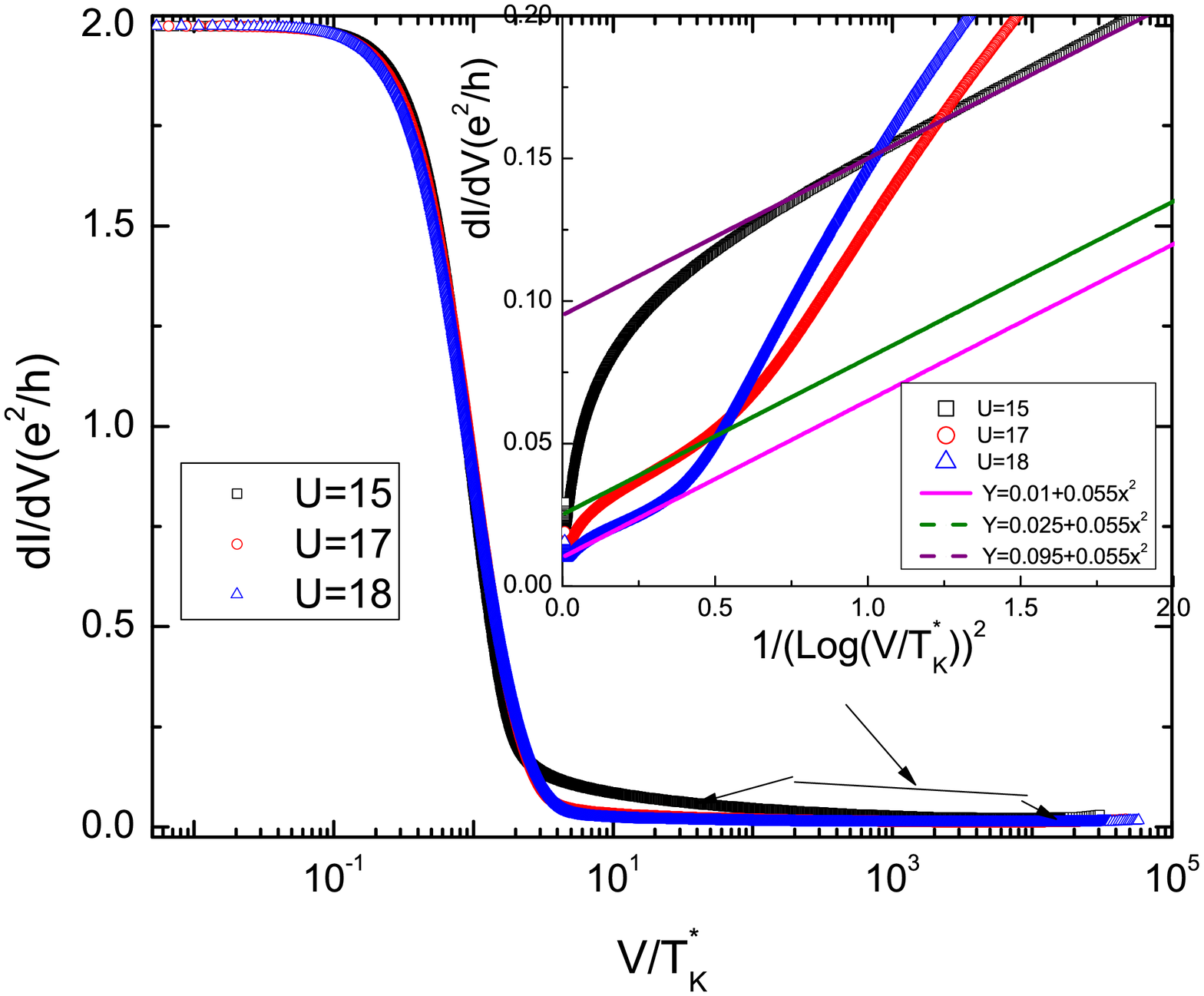}
	\caption{Top: Zoomed in picture of the differential conductance vs voltage nearby zero voltage. Inset shows
        the	universality in conductance vs voltage scaled by $T_k^*$ when $\frac{V}{T_k^*}\leq 1$.
	  The quadratic behavior occurs for $\frac{V}{T_k^*}<0.5$ as indicated by the fitted curve. Bottom: Differential conductance vs voltage scaled by $T_k^*$
	  nearby the Kondo peak structure. Inset shows the logarithmic behavior when $\frac{V}{T_k^*}\gg 1$. $\Gamma=0.5$ for all these data sets.
  	}\label{didvneqfig3}
\end{figure}

\begin{figure}[h]
	\includegraphics[width=1\columnwidth, clip]{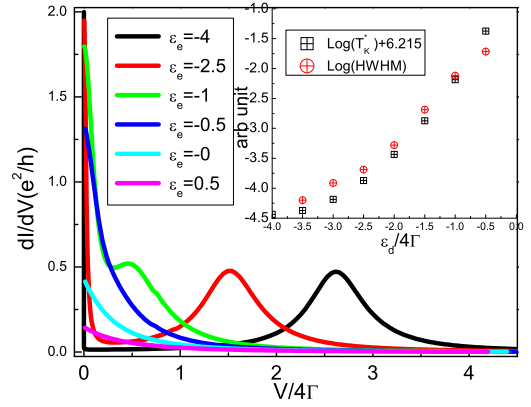}
	\caption{
	 	$dI/dV$ vs $V/4\Gamma$ for $U=8$, $\Gamma=0.25$ and
	  various $\epsilon_d$ from Kondo ($\epsilon_d=-4$) to mixed valence regime ($\epsilon_d\simeq 0$).
	  Inset: Comparison of $\ln(T_k^*)-\ln(c_1)$ and $\ln(V_{HWHM})$ as a function of impurity level $\epsilon_d$.
	  Here $V_{HWHM}$ is the voltage difference estimated at half value of differential conductance at zero voltage.
      The constant shift $-\ln(c_1)$ is chosen to give the best fit in the data away from $\epsilon_d=-\frac{U}{2}$.
  	}\label{didvneqfig2}
\end{figure}

	The plots of the differential conductance vs source drain voltage
	for different dot levels, $\epsilon_d$, tunneling strengths $\Gamma$
	and interaction strengths $U$ are shown in Fig.~\ref{didvneqfig1} and Fig.~\ref{didvneqfig2}.
	Two major features emerge from these plots: 1) A narrow peak around zero bias
	reaching maximal value of $2e^2/h$ (the unitary limit) for values of the
	gate voltage close to the symmetric point ($\epsilon_d \simeq -U/2$).
	2) A broader peak developing at finite bias. The first
	peak is a non-perturbative effect identified as the many body Kondo peak, characteristic of
	strong spin fluctuations in the system. But the broad peak
	is due to renormalized charge fluctuations around the impurity
	level. Notice the two features merge as the gate voltage,
	$\epsilon_d$ is raised from the Kondo regime, $\epsilon_d=-U/2$, to
	the mixed valence regime, $\epsilon_d=0$, with the Kondo effect
	disappearing. As a function of the bias the various curves
	describing the Kondo peak for different values of the parameters can
	be collapsed onto a single universal function
	$\mathrm{d}I/\mathrm{d}V= \mathrm{d}I/\mathrm{d}V(V/T_k^*)$ as shown in
	Fig.~\ref{didvneqfig3}. Here $T_k^*$ is defined as
	\begin{equation}
	\label{TKstar}
	T_k^*=c_1\frac{\sqrt{2U\Gamma}}{\pi}e^{\frac{\epsilon_d(\epsilon_d+U)+\Gamma^2}
	{2U\Gamma}}
	\end{equation}
	with $c_1=0.002$. The energy scale $T_k^*$ was extracted from the numerics by requiring that
	the function $\mathrm{d}I/\mathrm{d}V(V/T_k^*)$ decreases to half its maximal value when
	$V\simeq T_k^*$. The expression for $T_{k}^\ast$ as given by Eq.~(\ref{TKstar}) differs from
	the thermodynamic $T_k$ as defined in Eq.~(\ref{TK}). The difference of prefactor in the exponential is certainly related to the unusual choice of regularization scheme in the SBA~\cite{AFL}.
	The other possible implication for this different formulation for the Kondo scale is also
	addressed later when we discuss the experiment done by L. Kouwenhoven
	et al\cite{WG}.\\

	The small voltage behavior for differential conductance in symmetric case, i.e.
	$\epsilon_d\simeq-\frac{U}{2}$, is expected to be~\cite{Oguri2,Glazman}
	$$
	\frac{dI}{dV}\Big|_{V\ll T_k^*}\simeq \frac{2e^2}{h}\left(1-\alpha_V\left(\frac{V}{T_k^*}
	\right)^2\right)
	$$
	and allows us to identify the constant $\alpha_V$ from the quadratic deviation from $2e^2/h$.
	The quadratic fit of the universal curve around $V\simeq 0$, as shown in
	Fig.~\ref{didvneqfig3}, gives
	$\alpha_V\simeq 1$.
	It is also expected for $T_k^*\ll V\ll \frac{U}{2}$ that the tail of the peak
	decays logarithmically ~\cite{Glazman} as
	$$\frac{dI}{dV}\sim \frac{2e^2}{h} \frac{1}{\ln^2(\frac{V}{T_k^*})}\ .$$
	The latter
	behavior is observed (see inset of Fig.~\ref{didvneqfig3} )
	in the regime $\frac{U}{\Gamma}\gg1$ for $10^2<\frac{V}{T_k^*}<10^4$ with the logarithmic function
	given by
	$$\frac{dI}{dV} = \frac{e^2}{h}\left[f\left(\frac{U}{\Gamma}\right)
	+\frac{c_2}{\ln^2(\frac{V}{T_k^*})}\right]$$
	with the parameter $c_2=0.055$. Here $f(\frac{U}{\Gamma})$
	is simply a constant (in $V$) shift. As suggested from the bottom plot of Fig.~\ref{didvneqfig3} (see also Fig.~\ref{infdidv} for the infinite $U$ case) the charge fluctuation side peak does not fall into
	the same scaling relation but the strong correlations shift the center of the side peak closer to $V=0$
	(see Fig.~\ref{didvneqfig1} and Fig.~\ref{didvneqfig2}). In other words the position of the
	 resonance in the $dI/dV$ curve naively expected around $V=|\epsilon_d|$ is
	 renormalized~\cite{Haldane} by the presence of interactions.
	In the inset of Fig.~\ref{didvneqfig2} we show the logarithm of the voltage obtained at half width
	half maximum (HWHM) of
	the zero voltage peak and compare it with
	$$\ln T_k^* =\frac{\epsilon_d(\epsilon_d+U)+\Gamma^2}{2U\Gamma}
	+\ln\left(c_1\frac{\sqrt{2U\Gamma}}{\pi}\right)$$
	(after subtracting the constant $\ln c_1$). What is important and \emph{universal} is that both quantities
	($\ln V_{\mathrm{HWHM}}$ and $\ln T_k^*$) exhibit a quadratic behavior in the gate voltage
	$\epsilon_d$. Similar results
	had been found experimentally by L. Kouwenhoven
	et al~\cite{WG} when they compare the full width half maximum of $dI/dV$ (from which they obtain a Kondo scale $T_{k_1}$ at finite voltage) with the temperature dependence of the linear
response differential conductance (from which another Kondo scale $T_{k_2}$ is extracted).
It is suggested from our numerical results that
both $\ln T_{k_2}$ (in analogy with our $T_k$) and $\ln T_{k_1}$ (which is our $T_k^*$) follows similar quadratic behavior in $\epsilon_d$ but differ in their curvatures by a factor of $\pi$. In Ref.~\onlinecite{WG} the curvatures of the quadratic behavior differ by a factor of around $2$ (see Fig.3B in Ref.~\onlinecite{WG}) which is attributed to dephasing of spin fluctuations at finite voltage.

	Notice that in all the numerical data shown for current vs voltage we have chosen
	$\frac{U}{\Gamma}\geq8$ to explore the scaling relation in the Kondo regime. Another reason is that
  our \emph{phenomenological distribution functions} introduced to
	control the relative weight for spin- and charge-fluctuation
	contributions work is much better in the large $\frac{U}{\Gamma}$ regime
	(cf. Fig.~\ref{eqdgplot}).

\begin{figure}[h]
	\includegraphics[width=1\columnwidth, clip]{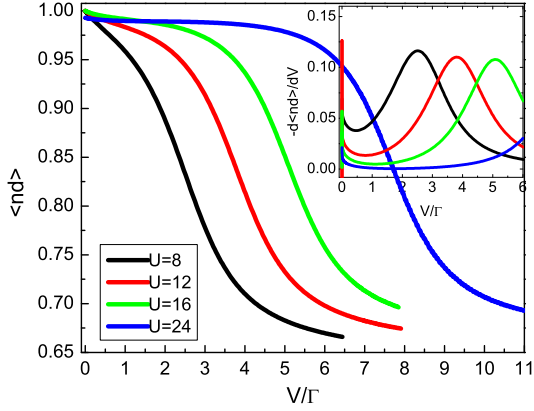}
	\caption{$\langle\hat{n_d}\rangle$ vs $V/\Gamma$ for different $U$ with $\epsilon_d=-\frac{U}{2}$ and $\Gamma=1$ case.
	  Inset: The corresponding nonequilibrium charge susceptibility. A small peak shows up nearby $V=0$ for all these curves.}\label{didvneqfig4}
\end{figure}

\begin{figure}[h]
	\includegraphics[width=1\columnwidth, clip]{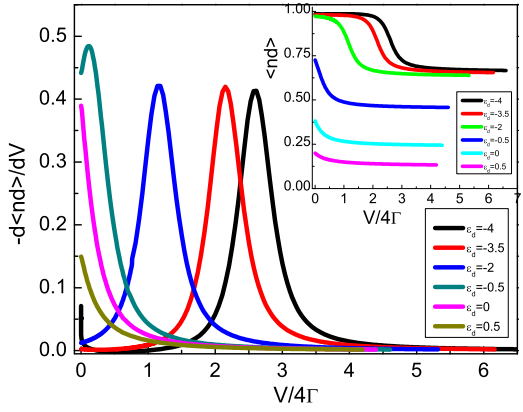}
	\caption{$-\frac{d\langle\hat{n}_d\rangle}{dV}$ vs $V/4\Gamma$ for $\Gamma=0.25$, $U=8$, and various $\epsilon_d$
	 from Kondo to mixed valence regime. We see that the small peak nearby $V=0$ only appears when $\epsilon_d
	 \rightarrow-\frac{U}{2}$. Inset: The corresponding $\langle\hat{n_d}\rangle$ vs $V/4\Gamma$.}\label{didvneqfig5}
\end{figure}

\begin{figure}[h]
	\includegraphics[width=1\columnwidth, clip]{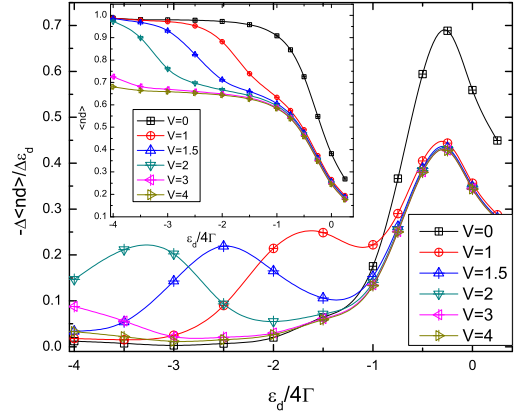}
	\caption{$-\frac{\Delta\langle\hat{n}_d\rangle}{\Delta\epsilon_d}$ for various fixed voltages as a function of
	 $\epsilon_d$ for $\Gamma=0.25$, $U=8$. Inset shows $\langle\hat{n}_d\rangle$ vs $\epsilon_d$ for various fixed voltage.}\label{didvneqfig6}
\end{figure}

 	Next let us study the change in the dot occupation as a function of the voltage.
 	The extension of the computation of the dot occupation out of equilibrium is straightforward.
	Suppose we find the correct distribution functions $f_s(\lambda)$ and $f_h(\lambda)$ then we have $\nu^{SBA}(\lambda)=\nu^s(\lambda) f_s(\lambda)+\nu^h(\lambda)f_h(\lambda)$. Under this assumption $\nu^{SBA}(\lambda)$
	retains its form
	in and out-of-equilibrium and the general expression for $\langle\hat{n}_d\rangle$ is
	\begin{eqnarray}
	\label{dot}
	&&n_d(\mu_1,\mu_2) =  \langle\Psi, \mu_1,\mu_2|\hat{n}_d|\Psi,
	\mu_1,\mu_2\rangle \\
	&&=2\Bigg(\int_{B_{1}}^{\infty}\mathrm{d}\lambda\ \sigma_b(\lambda)
	\nu^{SBA}(\lambda)+\int_{B_{2}}^{\infty}\mathrm{d}\lambda\ \sigma_b(\lambda)
	\nu^{SBA}(\lambda)\Bigg)\nonumber
	\end{eqnarray}
	As the form for $\nu^{SBA}(\lambda)$ is proved to be
	\emph{exact}  
	in equilibrium, we shall regard Eq.~(\ref{dot}) as an \emph{exact}
	result for $\langle\hat{n_d}\rangle$ even out of equilibrium and valid in all different 
	range of $U$,
	$\epsilon_d$, $\Gamma$ under the \emph{assumption} that the integrand does not change its form for in and out of equilibrium, which is the case for general
results of SBA. In the numerical results shown hereafter we shall use this
	expression, Eq.(\ref{dot}), for matrix element of dot occupation rather than Eq.~(\ref{condition}). We adopt the same
	voltage drive scheme by fixing $\mu_1$ and lowering $\mu_2$.
 	 	
	By using this result we do not need to confine ourself for large $\frac{U}{\Gamma}$.
	The case for different $\frac{U}{\Gamma}$ with $\epsilon_d=-\frac{U}{2}$ and for $U=8,\Gamma=0.25$
	with different $\epsilon_d$ are shown in Fig.~\ref{didvneqfig4} and Fig.~\ref{didvneqfig5}.
	The main features of these plots are a relatively slow decrease of the dot occupation at low
voltage followed by an abrupt drop of $\langle n_d\rangle$. The decrease of $\langle n_d \rangle$ takes place within a range of voltage of the order of $\Gamma$. Then as we increase the voltage
further another plateau develops. Note that, as expected, the bigger $U$ is the higher the voltage needed to drive the system out of the $\langle n_d\rangle =1$ plateau. In a sense the charge fluctuations are strongly frozen at large $U$ and it costs more energy to excite them. The voltage where the abrupt drop in
 $\langle n_d\rangle$ occurs corresponds to the energy scale at which the "charge fluctuation peak" was
 observed in the conductance plots. This can be seen by comparing the position of the broader peak
  in Fig.~\ref{didvneqfig2} with that of the abrupt dot occupation drop in Fig.~\ref{didvneqfig5}.
	
	Similar to the differential
	 conductance we may define the \emph{nonequilibrium charge susceptibility} as
	 $$\chi_c(V)|_{\epsilon_d}=-\frac{\partial\langle \hat{n_d}\rangle}{\partial V}\ $$
	 that we obtain by taking a numerical derivative of the dot occupation data with respect to
	 the voltage.
	 In the case of $U=-\epsilon_d/2$ there are two features as can be seen from the inset of
	 Fig.~\ref{didvneqfig4} and main figure of Fig.~\ref{didvneqfig5}. Nearby $V\simeq0$ we see a
	 first small peak arising with width and height decreasing
	 with increasing $\frac{U}{\Gamma}$. We identify this peak as a small remnant of the
	 charge fluctuations in the Kondo regime. This statement is confirmed by noticing that
	  this peak goes away as $\frac{U}{\Gamma}$ increases, vanishing when $U\rightarrow\infty$ as
	  shown in Section III where the infinite $U$ Anderson model is discussed.
	  The second peak is located at the same voltage as the charge
	  fluctuation peak observed in the conductance plots and is therefore associated to the
	  response of the renormalized impurity level to the charge susceptibility. This can be
	  seen when comparing Fig.~\ref{didvneqfig2} and Fig.~\ref{didvneqfig5}.
	
	  Another interesting quantity, the usual \emph{charge susceptibility}, defined by
	  $\chi_c(\epsilon_d)|_{V}=-\frac{\partial\langle \hat{n_d}\rangle}{\partial \epsilon_d}$, can also be
	  qualitatively described. In Fig.~\ref{didvneqfig6} we plot $-\frac{\Delta
	  \langle \hat{n_d}\rangle}{\Delta \epsilon_d}$ as a function
	  of $\epsilon_d$
	as we only have a few points in fixed $\epsilon_d$ for finite voltage. Notice that
	$\chi_c(\epsilon_d)|_{V}$ tends to be an universal
	curve in large voltage, indicating charge on the dot remains at some constant value in the steady
	 state with large voltage. This constant value at large voltage, as pointed out by C. J. Bolech, is
	 around $0.65$ for $\epsilon_d=-\frac{U}{2}$ case. In preparing this article we noticed that a
	 similar computation, adopting the same asymmetric voltage drive protocol as we have here,
is carried out by R. V. Roermund et al~\cite{Rapha} for the dot occupation out of equilibrium by using
equation of motion method. We do get a similar value for the dot occupation at large voltage.
This value is
different from the dot occupation value $n_d\simeq0.5$ at large voltage when the interaction $U$ is turned
off as shown in Fig.~\ref{infndv}. This difference might have to do with the $0.7$ structure observed in quantum
point contact\cite{SM} in high temperature (temperature is high compared with the Kondo scale but still
small compared with phonon modes or electronic level) and zero magnetic field as the linear response
conductance given by $n_d=0.65$ by using Friedel sum rule is around $0.73$. In a sense the voltage
seems to play a similar role to the temperature on the way it influences the dot occupation. Further connection
between these two behaviors could be clarified by computing the decoherence factor as in Ref.
~\onlinecite{Rapha}. This decoherence factor is related to the dot correlation function out of equilibrium
which can be computed in three-lead setup~\cite{Eran} by using our approach. 	
\subsection{Comparison with other theoretical and experimental results}
\begin{figure}[h]
	\includegraphics[width=1\columnwidth,height=60mm, clip]{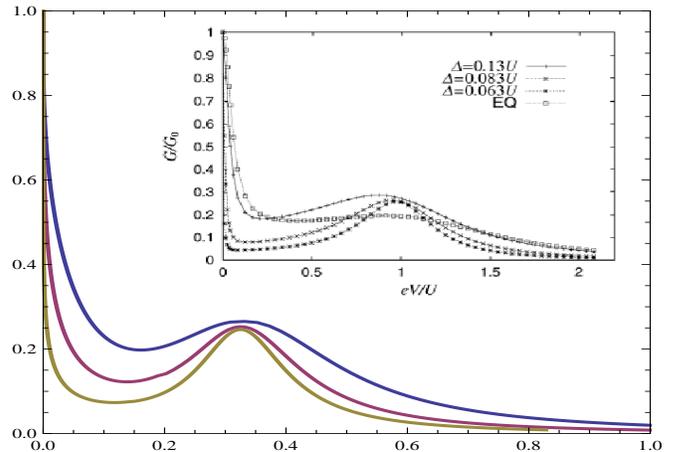}
	\caption{Comparison of our theory with perturbation expansion in $U$ done by D. Matsumoto on $dI/dV$ (y-axis in unit of $2e^2/h$) vs $V/U$ (x-axis). Our data (Blue, purple, and brown lines correspond to $\frac{\Gamma}{U}=0.13,0.083,0.063$ respectively. $\Delta$ shown in inset is $\Gamma$ in our notation. EQ in the inset is conductance computed by equilibrium density of state which is not relevant to our discussion here.) is shown as the main figure and Fig.8 in Ref~\onlinecite{Daichi} is shown in the inset. In Ref~\onlinecite{Daichi} the voltage is driven symmetrically, i.e. $\mu_1=-\mu_2$, rendering the factor of two difference in the voltage (i.e. $\frac{V}{U}=0.5$ in our case corresponds to $\frac{eV}{U}=1$ in the inset. $e=1$ in our convention.) in comparing our result with that in Ref~\onlinecite{Daichi}.}\label{cth}
\end{figure}

In most of the other theoretical approaches~\cite{Rosch,Eckel,konik2,konik,Rapha,Daichi,Dagotto} the symmetric voltage drive ($\mu_1=-\mu_2$) is usually assumed to preserve particle-hole symmetry in symmetric case ($\epsilon_d=-\frac{U}{2}$). It is thus
difficult for us to make any definite comparison with other theoretical results. The qualitative feature, as
shown by the black curves in Fig.~\ref{cth} done by D. Matsumoto~\cite{Daichi} by using perturbation
expansion in $U$ at strong coupling fixed point, is similar to our results in the sense that the height of
the charge fluctuation side peak and width are almost the same. The major differences are in the
shape of Kondo peak and the position of the charge fluctuation side peak. A clear signature of
renormalized dot level $\epsilon_d$ as hinted in renormalization computation~\cite{Hewson,Haldane}
is clearly seen in our result. The shape of Kondo resonance nearby zero voltage deviates from its
 quadratic behavior expected from Fermi liquid picture at smaller voltage in our case as is expected
for asymmetric voltage drive~\cite{Oguri,Zurab}.
\begin{figure}[h]
	\includegraphics[width=.9\columnwidth, clip]{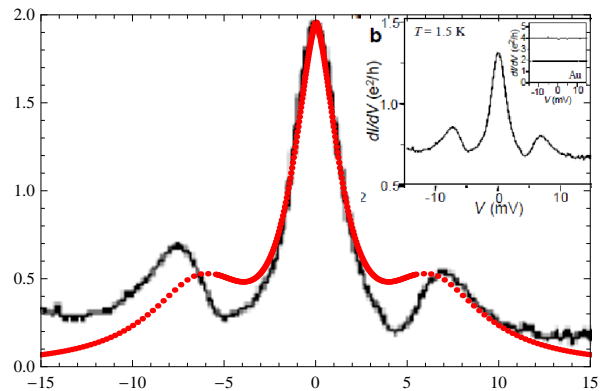}
	\caption{Comparison of theory with experiment of $dI/dV$ (y-axis in unit of $e^2/h$) vs $V$ (x-axis in unit of $mV$). Inset is the original data graph published in Ref.~\onlinecite{Park}. The red dots are given by our theory for $\frac{U}{\Gamma}=8$ with voltage rescaled to fit with original data in unit of $mV$. The value of differential conductance (experiment data in black line) is rescaled from $(0.6,1.3)$ to $(0,2)$ in unit of $\frac{e^2}{h}$.}\label{cexp}
\end{figure}

We can also compare our results with experiments. As shown in the inset of Fig.~\ref{cexp} is the
$\frac{dI}{dV}$ vs $V$ measured in Co ion transistor by J. Park et al. \cite{Park}. We rescaled the differential conductance and superimposed our numerical results on the data graph. The measurement was done by using an asymmetric drive of the voltage (by keeping $\mu_1=0$ and changing $\mu_2$ to be larger or smaller than zero) and thus there is an asymmetry in the differential conductance as a function of voltage as illustrated in the data curve. In our numerics we only compute the scenario for $\mu_1=0$ and lowering $\mu_2$ (only
for $V>0$ region of Fig.~\ref{cexp}). The $V<0$ region is plotted by just a reflection with respect to the
 $V=0$ axis which illustrates the case of $\mu_2=0$ and lowering $\mu_1$. To compare with the correct voltage setup on the $V<0$ side as in experiment will involve computations within a different parametrization for bare the Bethe momenta
which is beyond our current scope. The comparison on the $V>0$ region shows good agreement between our theory and experimental result. The discrepancy
on the width of the charge fluctuation side peak could be due to the vibron mode~\cite{Jens}. To describe these type of transistors we shall start with the Anderson-Holstein Hamiltonian. We are currently exploring the possibility of solving this model by the Bethe Ansatz approach.
\section{Infinite $U$ Anderson model}
\label{C}
In the limit of $\frac{U}{\Gamma}\rightarrow\infty$ the finite $U$ two-lead Anderson impurity Hamiltonian becomes
the two-lead infinite $U$ Anderson model. The latter model is closely related, via the Schrieffer-Wolff transformation\cite{HewsonBook}, to the notorious Kondo model, a model of spin coupled to a Fermi liquid bath. The reason for that is simple: since $U\to\infty$ the charge fluctuations are essentially frozen out and only the spin fluctuations dominate the low-energy physics. The Hamiltonian is given by
\begin{multline}
	\hat{H}=\sum_{i=1,2}\int \dd x\,
	\psi^{\dagger}_{i\sigma}(x)(-i\partial_x)\psi_{i\sigma}(x)+
	\epsilon_d d^{\dagger}_\sigma
	d_\sigma\\
	+t_i(\psi^{\dagger}_{i\sigma}(0)b^{\dagger} d_\sigma+d^{\dagger}_
	\sigma b\psi_{i\sigma}(0))
	\label{infUAn}
\end{multline}
Here the bosonic operator $b$ is introduced to conserve $b^{\dagger}b+\sum_{\sigma}d^{\dagger}_{\sigma}d_{\sigma}=1$ and by applying the slave boson technique we project out
the phase space of double occupancy occurring in finite $U$ case.
The corresponding Bethe momenta distribution function for the infinite $U$ Anderson model is given by
\begin{multline}
2\sigma(\Lambda)=\frac{1}{\pi}-\int_{-\infty}^{B_2} d\Lambda'K(\Lambda-\Lambda')\sigma(\Lambda')\\
-\int_{-\infty}^{B_1} d\Lambda'K(\Lambda-\Lambda')\sigma(\Lambda')
\label{infUdis}
\end{multline}
with $K(\Lambda)=\frac{1}{\pi}\frac{2\Gamma}{(2\Gamma)^2+(\Lambda-\Lambda')^2}$.

\begin{figure}[h]
	\includegraphics[width=1\columnwidth, clip]{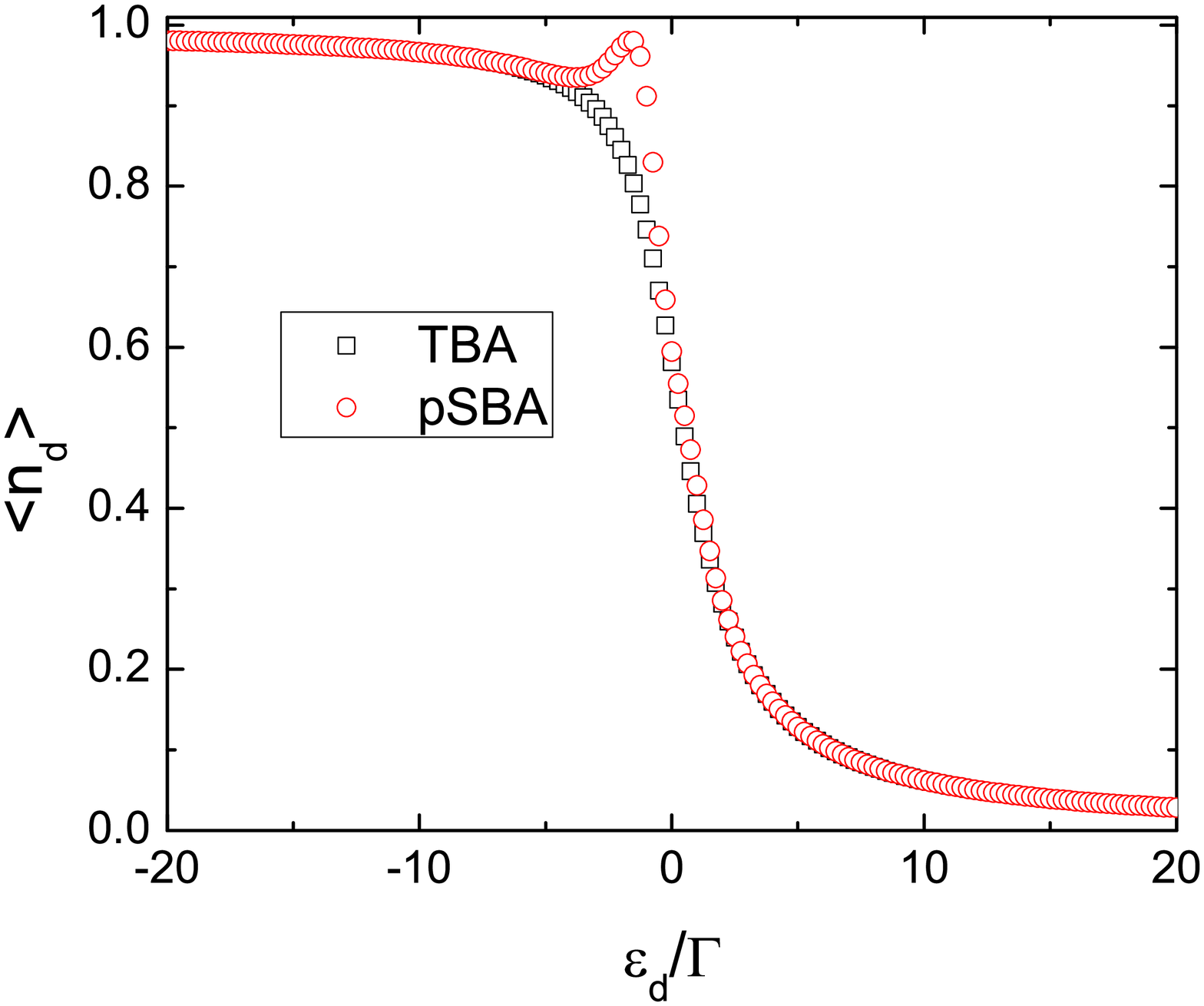}\\
	\includegraphics[width=1\columnwidth, clip]{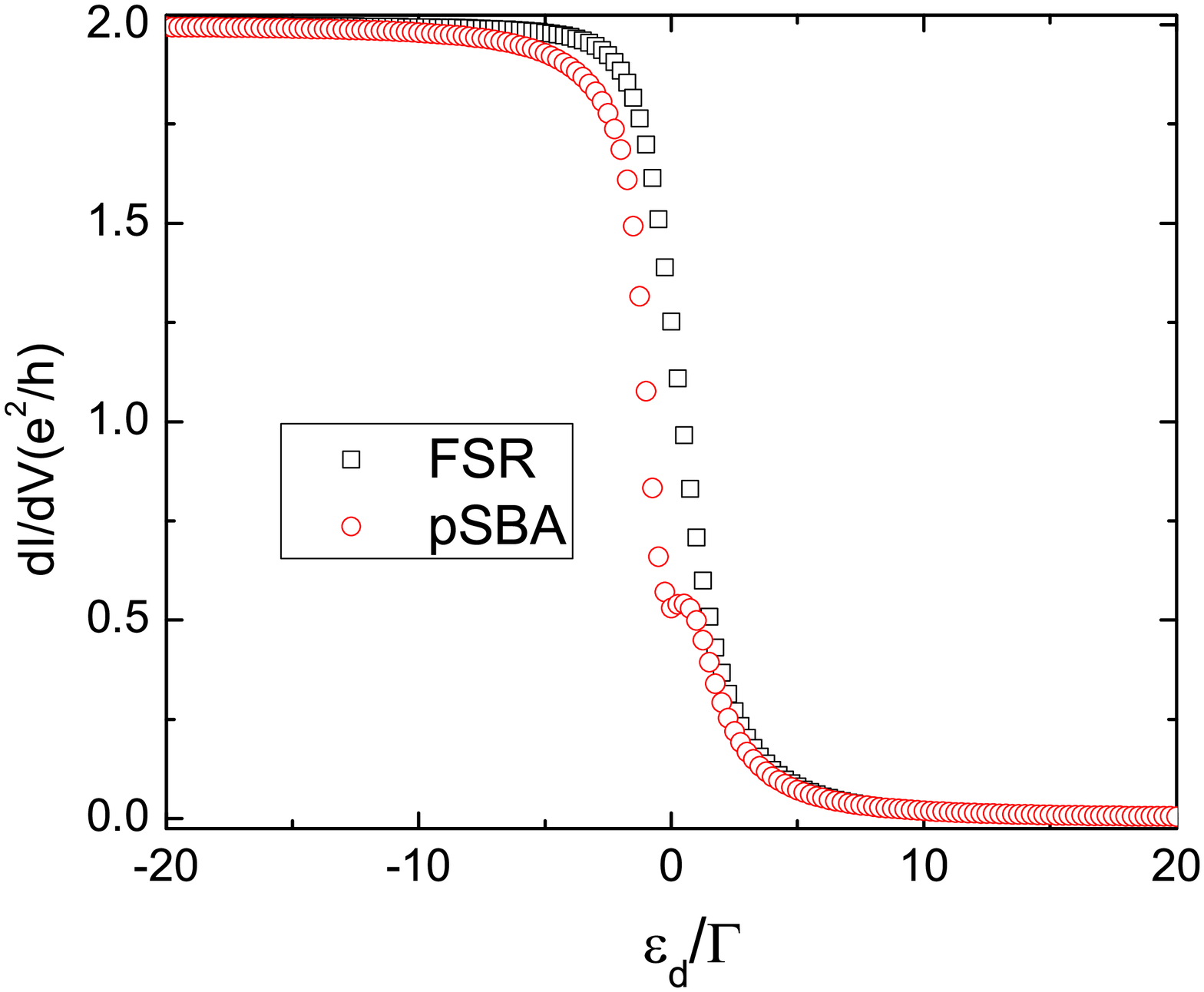}
	\caption{Top: $\langle \hat{n}_d\rangle$ vs $\frac{\epsilon_d}{\Gamma}$ for exact TBA result and pSBA. Bottom: Linear response conductance $dI/dV|_{V\rightarrow 0}$ vs $\frac{\epsilon_d}{\Gamma}$ for exact result (FSR) and pSBA in the infinite $U$ Anderson model. $\frac{D}{\Gamma}=-100$. Similar to the case of finite $U$ the
comparison nearby mixed valence region ($\epsilon_d\simeq 0$) is poorer.}\label{infnd}
\end{figure}

Eq.~(\ref{infUdis}) can be derived directly following the procedures in the finite $U$ Anderson model. It can also be derived from the finite $U$ result, Eq.~(\ref{bulk}), by taking the large $U$ limit ($U\gg\epsilon_d$, $U\gg\Gamma$):
\begin{eqnarray}\nonumber
\frac{x(\lambda)}{U}&\rightarrow&\frac{1}{2}-
\sqrt{\frac{\frac{\lambda}{U^2}+\frac{1}{4}+\sqrt{(\frac{\lambda}{U^2}+\frac{1}{4})^2+\frac{\Gamma^2}{U^2}}}{2}}\\
&\rightarrow&\frac{1}{2}-\sqrt{\frac{\frac{\lambda}{U^2}+\frac{1}{4}+|\frac{\lambda}{U^2}+\frac{1}{4}|}{2}}\\\nonumber
&\rightarrow&\frac{1}{2}-\frac{1}{2}(1+\frac{2\lambda}{U^2}+\ldots)\rightarrow-\frac{\lambda}{U^2}=\frac{\Lambda}{U}\\\nonumber
\frac{y(\lambda)}{U}&\rightarrow&\sqrt{\frac{-(\frac{\lambda}{U^2}+\frac{1}{4})+((\frac{\lambda}{U^2}+\frac{1}{4})^2+\frac{\Gamma^2}{U^2})^{1/2}}{2}}\\&\rightarrow&\sqrt{\frac{(\frac{\lambda}{U^2}+\frac{1}{4})(-1+(1+\frac{(\frac{\Gamma}{U})^2}{(\frac{\lambda}{U^2}+\frac{1}{4})^2})^{1/2})}{2}}\\\nonumber
&\rightarrow&
\left(\frac{1}{4}\frac{(\frac{\Gamma}{U})^2}{\frac{1}{4}}\right)^{1/2}+\mathcal{O}(U^{-2})\simeq\frac{\Gamma}{U}
\end{eqnarray}
with $\Lambda\equiv-\frac{\lambda}{U}$. Similar procedures as in Appendix \ref{B} give the matrix element $\nu^{SBA}_{\infty}(\Lambda)$ for the dot occupation in the infinite $U$ Anderson model in equilibrium to be
\begin{eqnarray}
\nu^{SBA}_{\infty}(\Lambda)=\frac{2\Gamma}{(\Lambda-\epsilon_d)^2+(2\Gamma)^2}\ .
\end{eqnarray}
In going to the out-of-equilibrium regime ($\mu_1\neq\mu_2$) we follow the same phenomenological method as for the finite $U$ case.
The result for the spin-fluctuation and charge-fluctuation contributions to the dot occupation are given by
\begin{eqnarray}\nonumber
&&\nu_{\infty}^s(\Lambda)=\frac{1}{\Gamma}\left(1-\frac{\epsilon_d-\Lambda}{\sqrt{(\epsilon_d-\Lambda)^2+4\Gamma^2}}\right)\\
&&\nu_{\infty}^h(\Lambda)=\frac{2\Gamma}{(\Lambda-\epsilon_d)^2+(2\Gamma)^2}\ .
\end{eqnarray}
We shall again check the consistency with the exact result for the dot occupation in equilibrium,
 namely
\begin{eqnarray}\nonumber
&&\langle\sum_\sigma d^\dagger_\sigma d_\sigma\rangle=4\int_{D}^{B}\dd\Lambda\ \sigma_b(\Lambda) \nu_{\infty}^{SBA}(\Lambda)\\\nonumber
&&=4\int_{D}^{B}\dd\Lambda\ \sigma_b(\Lambda)( \nu_\infty^s(\Lambda)f_s^{\infty}(\Lambda)+\nu_\infty^h(\Lambda)f_h^{\infty}(\Lambda))\ .
\end{eqnarray}
Here $D$ is related to the bandwidth and $B$ is determined by the equilibrium Fermi energy $\mu_1=\mu_2=0$. $f_s^{\infty}(\Lambda)$ and $f_h^{\infty}(\Lambda)$
are expressed as
\begin{eqnarray}\nonumber
&&f_s^{\infty}(\Lambda)=\frac{T_k^{\infty}/\pi}{(\Lambda-B)^2+(T_k^{\infty})^2}\\\nonumber
&&f_h^{\infty}(\Lambda)=\frac{2\Gamma}{(\Lambda-B-\epsilon_d)^2+(2\Gamma)^2}\ .
\end{eqnarray}
Here the Kondo scale $T_k^{\infty}$ used in $f_s(\Lambda)$ takes the form\cite{Wing} $$T_k^{\infty}=\frac{\sqrt{10|D|\Gamma}}{\pi}e^{-\pi\frac{|\epsilon_d|}{\Gamma}}\ .$$
The results for the dot occupation and Friedel sum rule check in the infinite $U$ case are shown in Fig.\ref{infnd}. Again we see a nice match between
our phenomenological approach and the exact result for $|\frac{\epsilon_d}{\Gamma}|\neq 0$ and some mismatch in the mixed valence region$|\frac{\epsilon_d}{\Gamma}|\simeq 0$. This is consistent with the results for finite $U$.

\begin{figure}[h]
	\includegraphics[width=1\columnwidth, clip]{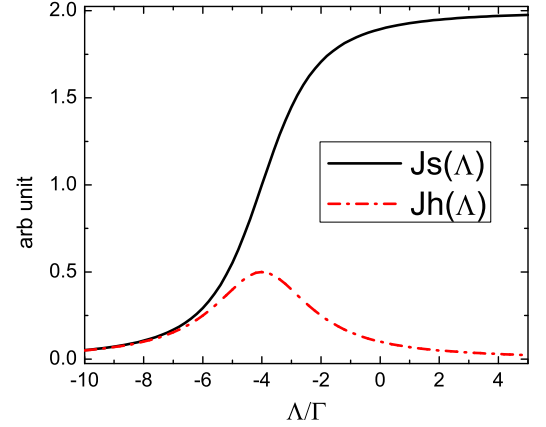}
	\caption{$J_s(\Lambda)$ and $J_h(\Lambda)$ vs Bethe momenta $\Lambda$ (scaled by $\Gamma$) in infinite U Anderson model. $\frac{\epsilon_d}{\Gamma}=-4$ in this graph. Similar graph appears for finite U case with x-axis replaced by real part of Bethe momenta $x(\lambda)$.}
	\label{mco}
\end{figure}
The corresponding spin and charge fluctuation matrix element for current, $J^s_{\infty}(\Lambda)$ and $J^h_{\infty}(\Lambda)$, are given by
\begin{eqnarray}\nonumber
&&J^s_{\infty}(\Lambda)=1-\frac{\epsilon_d-\Lambda}{\sqrt{(\epsilon_d-\Lambda)^2+4\Gamma^2}}\\\label{currentinf}
&&J^h_{\infty}(\Lambda)=\frac{2\Gamma^2}{(\Lambda-\epsilon_d)^2+(2\Gamma)^2}
\end{eqnarray}
The current expectation value is given by
\begin{eqnarray}\nonumber
\langle\hat{I}\rangle=\frac{2e}{\hbar}\int_{B_2}^{B_1}d\Lambda\sigma(\Lambda)
(J^s_{\infty}(\Lambda)f^{\infty}_s(\Lambda)+J^h_{\infty}(\Lambda)f^{\infty}_h(\Lambda))
\end{eqnarray}
where $B_1$ and $B_2$ are related to $\mu_1$ and $\mu_2$ by minimizing charge free energy $F$
\begin{eqnarray}\nonumber
F=2\left(\int_{D}^{B_1}\dd\Lambda\,  \sigma(\Lambda)(\Lambda-\mu_1)+\int_{D}^{B_2}\dd\Lambda\,\sigma(\Lambda)(\Lambda-\mu_2)\right)\ .
\end{eqnarray}

Before we proceed to discuss the numerical results for current vs voltage in this infinite $U$ model let us look at the
structure of $J^s_{\infty}(\Lambda)$ and $J^h_{\infty}(\Lambda)$ as a function of $\Lambda$ as shown in Fig.~\ref{mco}. $\Lambda$ here represents the bare
energy of the quasi-particle and plays the same role as $x(\lambda)$ in the finite $U$ Anderson model. $J^s_{\infty}(\Lambda)$
alone would reproduce the main feature in the Friedel sum rule for $\epsilon_d\ll 0$. In this region the linear response conductance comes mainly from the spin fluctuations. The upper plot of Fig.~\ref{mco} fixes $\epsilon_d$ and shows $J^s_{\infty}(\Lambda)$ vs $\Lambda$. We may also fix $\Lambda=0$ (in the sense
of choosing the equilibrium Fermi surface energy at $\Lambda=0$) and plot $J^s_{\infty}(\epsilon_d)$ vs $\epsilon_d$. In this way we can see that $J^s_{\infty}(\epsilon_d)$ vs $\epsilon_d$ reproduces the overall structure of the linear response conductance from the Kondo region ($\epsilon_d\leq 0$) to the mixed valence regime ($\epsilon_d\simeq 0$). Therefore
we identify the phase shift $\frac{\delta_{p^+}+ \delta_{p^-}}{2}$, contributing to $J^s_{\infty}(\Lambda)$, as the phase shift related to spin-fluctuation.

$J^h_{\infty}(\Lambda)$ gives a Lorentz shape in bare energy scale $\Lambda$. This structure is akin to the charge fluctuation side peak with peak position at
energy scale around $\epsilon_d$ as seen from lower plot of Fig.~\ref{mco}. Thus we identify the phase shift $\delta_{p^+}+ \delta_{p^-}$, contributing to $J^h_{\infty}(\Lambda)$, as the phase shift related to charge-fluctuation. These structures also apply to the case of the finite $U$ Anderson model.
\begin{figure}[h]
	\includegraphics[width=1\columnwidth, clip]{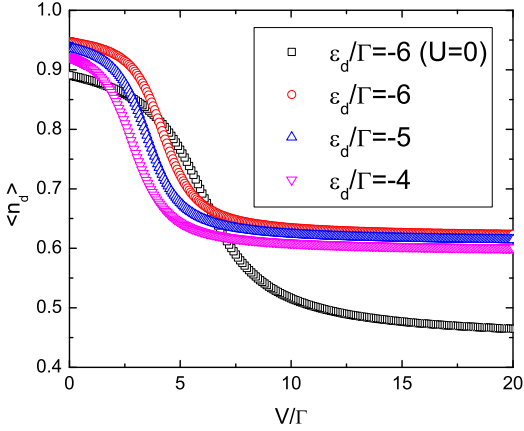}
	\caption{$\langle\hat{n}_d\rangle$ vs $\frac{V}{\Gamma}$ in infinite U Anderson model (for Red, Blue, and Purple dots corresponding to $\frac{\epsilon_d}{\Gamma}=-6,-5,-4$. The Black dots are $U=0$ and $\frac{\epsilon_d}{\Gamma}=-6$ case shown for comparison). $\frac{D}{\Gamma}=-100$ in this graph.}\label{infndv}
\end{figure}

Now let us discuss the out of equilibrium numerical results. The voltage is again driven asymmetrically by fixing $\mu_1\simeq0$ and lowering $\mu_2$. The \emph{exact} dot occupation vs voltage for different $\epsilon_d$ for infinite $U$ and $U=0$, $\frac{\epsilon_d}{\Gamma}=-6$ case (black dots) are shown
in Fig.~\ref{infndv}. We see again the dot occupation decreases slowly at low voltage and develops an abrupt drop at a voltage scale corresponding to impurity level $\epsilon_d$.
Also notice the apparent difference between the $U=0$ plot (black dots) and the
$U\rightarrow \infty$ case (red dots) and for the same value of $\frac{\epsilon_d}{\Gamma}$.
For $U\rightarrow \infty$, the dot occupation at large voltage is around $0.65$ for $\frac{\epsilon_d}{\Gamma}\ll 0$ which is consistent with the result of the finite $U$ case when $\frac{U}{\Gamma}$ is large (cf. Section II D). In contrast the non-interacting case ($U=0$) shows that
$\langle n_d \rangle\to 0.5$ at large bias.

\begin{figure}[h]
	\includegraphics[width=1\columnwidth, clip]{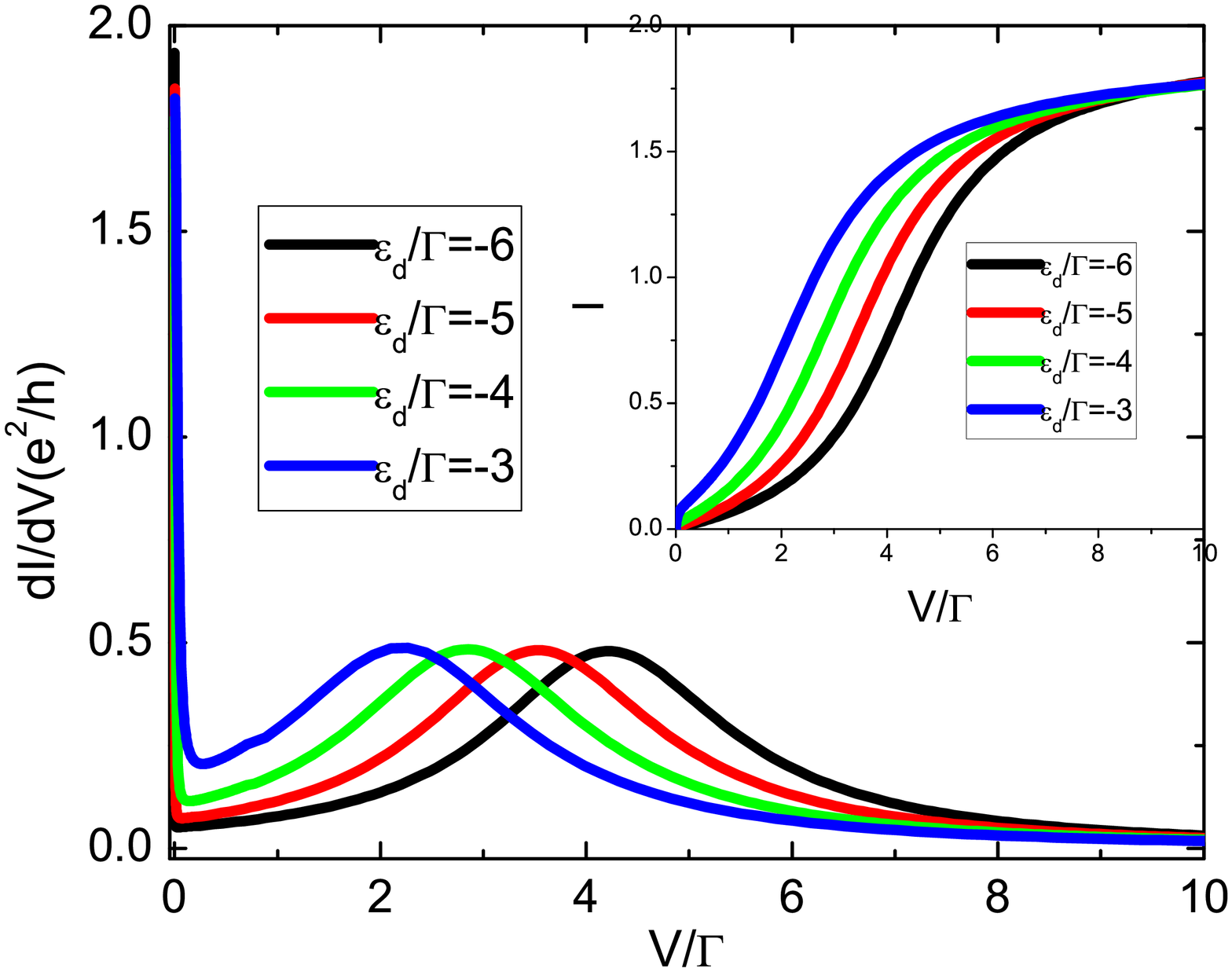}\\
	\includegraphics[width=.8\columnwidth, clip]{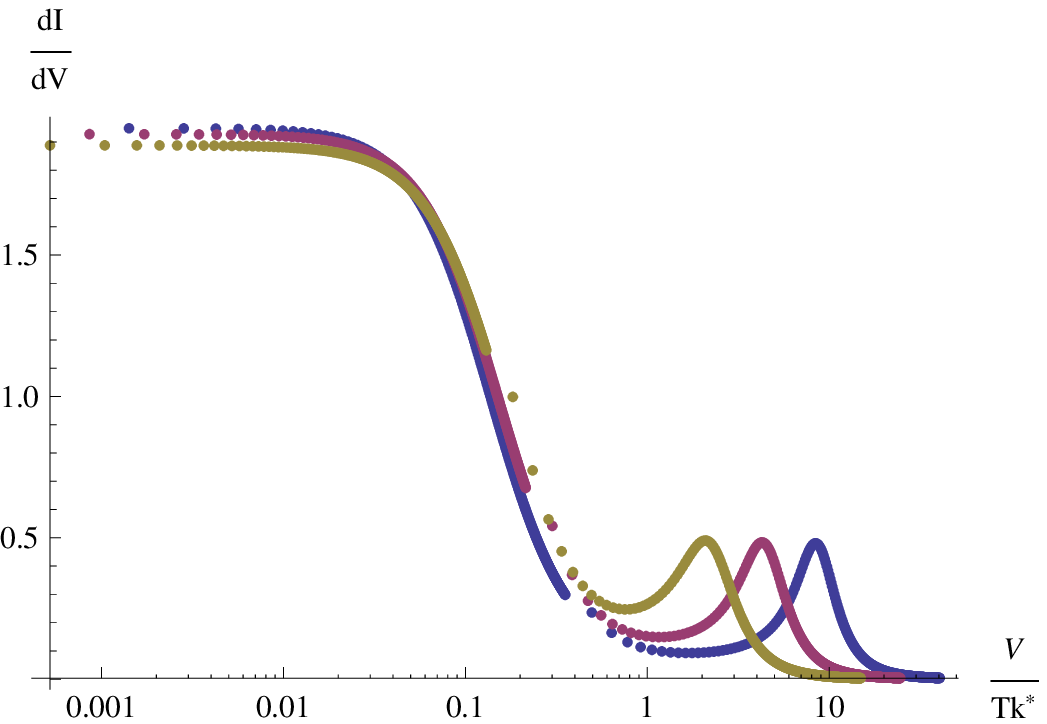}
	\caption{Top: $\frac{dI}{dV}$ vs $\frac{V}{\Gamma}$ in infinite U Anderson model. Inset shows the $I-V$ curves for these parameters. $\frac{D}{\Gamma}=-100$ in this graph. Bottom: $\frac{dI}{dV}$ vs $\frac{V}{T_k^*}$ shows the scaling relation nearby zero voltage for $\frac{\epsilon_d}{\Gamma}=-6,-5,-4$ (Blue, Purple, Brown).}\label{infdidv}
\end{figure}
The \emph{phenomenological} current vs voltage and the corresponding differential conductance vs voltage are plotted in the top figure of Fig.~\ref{infdidv}.
Again we see the zero bias anomaly and a broad charge fluctuation side peak in the differential conductance vs voltage.
The scaling relation of differential conductance vs voltage expected in small voltage region can also be extracted by rescaling the voltage by $T_k^{\infty *}$ as shown in bottom figure of Fig.~\ref{infdidv}. Here $T_k^{\infty *}$ is given by
$$T_k^{\infty *}=\frac{\sqrt{10|D|\Gamma}}{\pi}e^{-\pi\frac{|\epsilon_d|}{2\Gamma}}\ .$$
 Notice this $T_k^{\infty *}$ differs from $T_k^{\infty}$ with a factor of two within the exponent.
 This factor of two difference represents the difference in the curvature of the parabola as function of $\epsilon_d$ (the logarithm of half width at half maximum
 of the Kondo peak vs $\epsilon_d$ shows parabolic curve as in inset of Fig.~\ref{didvneqfig4} for finite U case). This factor of two ratio
bears even closer resemblance to the results shown in Ref.~\onlinecite{WG}.
Note that in bottom figure of Fig.~\ref{infdidv} the positions of the side peak are different and show no universality in that region. It shows universality for $\frac{V}{T_k^{\ast}}\leq 1$.
\section{Concluding Remarks}
 In this article we have explicitly computed the non-equilibrium
transport properties in the Anderson model for all voltages using
the Scattering Bethe Ansatz. In the case of equilibrium we have also shown the equivalence of traditional Bethe Ansatz
and Scattering Bethe Ansatz by evaluating dot occupation in equilibrium. For the expression of current we have
introduced \emph{phenomenological distribution functions} to set the weight for spin-fluctuation and charge-fluctuation contributions
to the current. The result shows correct
scaling relation in Kondo regime as well as satisfying
the Friedel sum rule for linear response for large $\frac{U}{\Gamma}$.

Other interesting quantities, such as the \emph{nonequilibrium charge susceptibility} or the usual charge susceptibility,
are computed numerically via \emph{exact} expression for dot occupation as
a function of voltage and impurity level. We believe this is the first report of an \emph{exact} computation of the dot occupation out-of-equilibrium and
it may have interesting application in quantum computing as we understand more the dephasing mechanism.
We have also compared our results with perturbation calculation and experimental measurement of nonlinear differential conductance of a quantum dot.

The major difficulty we encounter by using SBA comes from the single particle phase shift for complex momenta which leads to a breakdown of steady
state condition when out of equilibrium. One possible issue resulting in this is the local discontinuity at odd channel $s_{op}$, the choice
we made to enable us to construct a scattering state with fixed particles from lead 1 and lead 2. It can be proved that without this choice
we cannot write down fixed number of particles incoming from each lead\cite{foot1} in this Anderson impurity model and similarly for IRLM.
The other issue in the study for Anderson model is whether we shall include all possible bound states in
the ground state construction. From the mathematical structure we shall choose 4 type of bound states but the results from charge susceptibility seems to suggest 2 type of bound states is the correct choice. To check whether this is in general correct
we plan to come back to study the whole spectrum, which include bound state when Bethe energy higher
than impurity level, of IRLM as this model bares structure similarity to the Anderson model described in this article.
Following the SBA on IRLM~\cite{Mehta} there are lots of numerical approach and different exact methods~\cite{Boulat1} developed
for this model and detailed comparison for different approaches is desired for better understanding
its physics and scaling relation. By learning how to deal with complex momenta in this model we may also
find the rule which may lead us to the $\emph{exact}$ expression for current in this Anderson impurity model.
\section*{Acknowledgment}
We are grateful to Kshitij Wagh, Andres Jerez, Carlos Bolech, Pankaj Mehta, Avi Schiller, Kristian Haule, and Piers Coleman for many useful discussions and most particularly to Chuck-Hou Yee for his important help with the numerics and to Natan Andrei for numerous discussions and fruitful ideas. S. P. would also like to thank Daniel Ralph and Joshua Park for permission to use their data and discussion. G. P. acknowledges support from the Stichting voor Fundamenteel Onderzoek der Materie (FOM) in the Netherlands. This research
 was supported in part by NSF grant DMR-0605941 and DoEd GAANN fellowship.
\appendix
\section{Discussion of 2 strings vs 4 strings}\label{A}
As we have discussed in the main text the bounded pair, formed by $p^{\pm}(\lambda)=x(\lambda)\mp i y(\lambda)$,
can be formed by quasi-momenta from lead $1$ or lead $2$. We have shown the results for two type of strings (bound states). Namely the strings
are formed by $\{ij\}=\{11,22\}$ with $i$, $j$ denoting incoming lead indices. In this section we discuss the case of $4$ type of strings and show
thier corresponding numerical results in out of equilibrium regime (In equilibrium the $2$ strings and $4$ strings give the same result for dot occupation).

The density distribution for the Bethe momenta (rapidities) is denoted by $\sigma_{ij}(\lambda)$ with $\{ij\}=\{11,12,21,22\}$ indicating the incoming
	electrons from lead $i$ and lead $j$. The $\sigma_{ij}(\lambda)$ is given by
\begin{equation}
\label{4bulk} 4\sigma_{ij} (\lambda) =
-\frac{1}{\pi} \frac{\mathrm{d}x(\lambda)}{\mathrm{d}\lambda}
-\sum_{i,j=1,2}\int_{B_{ij}}^{\infty}\mathrm{d}\lambda'\
K(\lambda-\lambda')\sigma_{ij}(\lambda')
\end{equation}
The factor of $4$ indicates $4$ type of possible configurations and the constraint of exclusions in rapidities $\lambda$ in solving the quantum
inverse scattering problem. The idea is that in equilibrium four type of distributions are equally possible for each bound state bare energy $2x(\lambda)$.
The $B_{ij}$ play the role of chemical potentials for the Bethe-Ansatz
momenta and are determined from the physical chemical potentials of the two leads,
$\mu_i$, by minimizing the charge free energy,
\begin{equation}\nonumber
F=\sum_i (E_i-\mu_i N_i)=\sum_i \int_{B_{ij}}^{\infty}\mathrm{d}\lambda\
(x(\lambda)-\mu_i)\sigma_{(i)}(\lambda)d\lambda
\end{equation}
with $\sigma_{(1)}\equiv 2\sigma_{11}+\sigma_{12}+\sigma_{21}$ the
lead $1$ particle density and $\sigma_{(2)}\equiv
2\sigma_{22}+\sigma_{12}+\sigma_{21}$ the lead $2$ particle density.
In the case of $\mu_1\ > \mu_2$ we have $B_{11}<B_{12}=B_{21}<B_{22}$
for this finite U Anderson model but the equation for $\sigma_{ij} (\lambda)$
is the same for different combination of $i$ and $j$. The reason is we put a quasi-hole
state, rather than a quasi-particle, in the integral equation Eq.(\ref{4bulk}) similar to
the treatment of Wiener-Hopf approach. For example, for $B_{11} <\lambda < B_{22}$ there could be three type
of quasi-particle state $\{ij\}=\{11,12,21\}$ and we put $\{ij\}=\{22\}$ state as quasi-hole state. This hole state still
count one weight of the probability of $4$ distributions and therefore the factor of $4$ on the left hand side of Eq.(\ref{4bulk}) retains even out of equilibrium.
Similar idea is also applied in two type of bound state (strings) solution.

Other than their differences in the density distribution the
computations for the current and dot occupation expectation value are quite similar to the two strings case. We show their numerical results
in the following.

\begin{figure}[h]
	\includegraphics[width=1\columnwidth, clip]{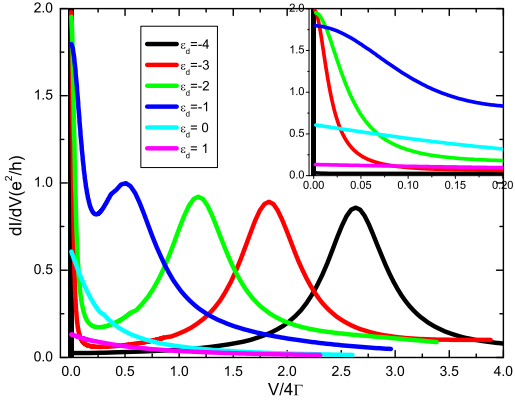}
	\caption{$\frac{dI}{dV}$ vs $\frac{V}{4\Gamma}$ for $U=8$, $\Gamma=0.25$ and various $\epsilon_d$ from $\epsilon_d=-\frac{U}{2}$ to $\epsilon_d=1$. The inset is the enlarged region nearby zero voltage.} \label{4sdidv}
\end{figure}

\begin{figure}[h]
	\includegraphics[width=1\columnwidth, clip]{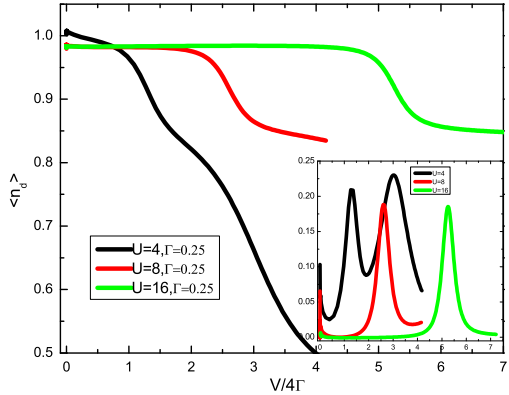}
	\caption{$\langle n_d\rangle$ vs $\frac{V}{4\Gamma}$ for different $U$, $\Gamma=0.25$ and $\epsilon_d=-\frac{U}{2}$. The inset is $-\frac{\partial\langle n_d\rangle}{\partial V}|_{\epsilon_d}$ vs $V$ voltage. A third peak shows up in $U=4$ case.}\label{4sndv}
\end{figure}

The differential conductance vs voltage as shown in Fig.\ref{4sdidv}, obtained by taking numerical derivative on current vs voltage data, essentially gives the same picture as in two strings
case, namely a sharp Kondo peak nearby $V=0$ and a broad side peak corresponding to charge fluctuations. In the case of $\langle n_d\rangle$ vs $V$, however, there
is an additional feature occurring at an energy scale higher than the energy scale of the charge fluctuation side peak (corresponding to the voltage position of $2$nd peak shown in the inset) as shown in Fig.~\ref{4sndv}. This is especially apparent if we looked at the nonequilibrium charge susceptibility as shown in inset of Fig.\ref{4sndv}.

As we do not expect there should be any
further charge fluctuations, we rule out, by physical argument, the possibility of $4$ strings configuration.
\section{Two particles solution and choice of $s_{op}$}\label{A-1}
For the two particles solution we follow similar construction in P B Wiegmann and A M Tsvelick's work\cite{PB} and the Scattering Bethe Ansatz approach developed by P. Mehta and N. Andrei\cite{Mehta}. Since Eq.(\ref{UAn}) is rotational invariant the spin quantum number is conserved. We show the solution with both particles with spin singlet incoming from lead 1 as an example in the following. Spin quantum number in z direction $S_z$ is a good quantum number and we can write the two particle solution of $S_z=0$ state as:
\begin{multline*}
|\Psi\rangle=\Big\{\int dx_1dx_2 \{ Ag(x_1,x_2)\psi^\dagger_{e\uparrow}(x_1)\psi^\dagger_{e\downarrow}(x_2)\\
+C h(x_1,x_2)\psi^\dagger_{o\uparrow}(x_1)\psi^\dagger_{o\downarrow}(x_2)
+B j(x_1,x_2)(\psi^\dagger_{e\uparrow}(x_1)\psi^\dagger_{o\downarrow}(x_2)\\
-\psi^\dagger_{e\downarrow}(x_1)\psi^\dagger_{o\uparrow}(x_2))\}+\int dx (A e(x)(\psi^\dagger_{e\uparrow}(x)d^\dagger_\downarrow-\psi^\dagger_{e\downarrow}(x)d^\dagger_\uparrow)\\
+Bo(x)(\psi^\dagger_{o\uparrow}(x)d^\dagger_\downarrow-\psi^\dagger_{o\downarrow}(x)d^\dagger_\uparrow))+Am d^\dagger_\uparrow d^\dagger_\downarrow \Big\}|0\rangle
\end{multline*}
Here $A,B,C$ are arbitrary constants to be determined later. To satisfy $\hat{H}|\Psi\rangle=E|\Psi\rangle=(k+p)|\Psi\rangle$ we have:
\begin{eqnarray}
\label{ee}
0 &=&[-i(\partial_{x_1}+\partial_{x_2})-E]g(x_1,x_2)\nonumber\\
&&+t[\delta(x_1)e(x_2)+\delta(x_2)e(x_1)]\\
\label{oo}
0&=&[-i(\partial_{x_1}+\partial_{x_2})-E]h(x_1,x_2) \\
\label{eo}
0 &=& [-i(\partial_{x_1}+\partial_{x_2})-E]j(x_1,x_2)+t\delta(x_1)o(x_2)\\
\label{ed}
0 &=& (-i\partial_x-E+\epsilon_d)e(x)+t g(0,x)+t \delta(x) m \\
\label{od}
0 &=& (-i\partial_x-E+\epsilon_d)o(x)+t j(0,x) \\
\label{dddd}
0 &=& (U+2\epsilon_d)m+2t e(0) -E m
\end{eqnarray}
For $U=0$ the model becomes non-interacting and the two particles solution becomes direct product of two one particle solutions.
\begin{multline*}
|\Psi\rangle=|\psi_{k\uparrow}\rangle\otimes|\psi_{p\downarrow}\rangle+|\psi_{p\uparrow}\rangle\otimes|\psi_{k\downarrow}\rangle\\
=\int dx_1dx_2 \{(g_k(x_1)\psi^\dagger_{e\uparrow}(x_1)+h_k(x_1)\psi^\dagger_{o\uparrow}(x_1)+e_k d^\dagger_\uparrow\delta(x_1))\\
(g_p(x_2)\psi^\dagger_{e\downarrow}(x_2)+h_p(x_2)\psi^\dagger_{o\downarrow}(x_2)+e_p d^\dagger_\downarrow\delta(x_2))\\
+(g_p(x_1)\psi^\dagger_{e\uparrow}(x_1)+h_p(x_1)\psi^\dagger_{o\uparrow}(x_1)+e_p d^\dagger_\uparrow\delta(x_1))\\
(g_k(x_2)\psi^\dagger_{e\downarrow}(x_2)+h_k(x_2)\psi^\dagger_{o\downarrow}(x_2)+e_k d^\dagger_\downarrow\delta(x_2))\}|0\rangle
\end{multline*}
Therefore at $U=0$ we have:
\begin{eqnarray}\nonumber
g(x_1,x_2)&=&g_k(x_1)g_p(x_2)+g_k(x_2)g_p(x_1)\\\nonumber
h(x_1,x_2)&=&h_k(x_1)h_p(x_2)+h_k(x_2)h_p(x_1)\\\nonumber
j(x_1,x_2)&=&g_k(x_1)h_p(x_2)+h_k(x_2)g_p(x_1)\\\nonumber
e(x)&=&e_k g_p(x)+e_p g_k(x)\\\nonumber
o(x)&=&e_k h_p(x)+e_p h_k(x)\\\nonumber
m&=&2e_p e_k
\end{eqnarray}
Now for $U\neq 0$ we shall derive the solution of this form
\begin{eqnarray}\nonumber
g(x_1,x_2)&=&Z_{kp}(x_1-x_2)g_k(x_1)g_p(x_2)\\&+&Z_{kp}(x_2-x_1)g_k(x_2)g_p(x_1)\label{g_ee}
\end{eqnarray}
Plug Eq.(\ref{g_ee}) into Eq.(\ref{ee}) we get
\begin{eqnarray}
e(x)=Z_{kp}(-x)g_p(x)e_k+Z_{kp}(x)g_k(x)e_p\label{e_ed}
\end{eqnarray}
Plugging above two results into Eq.(\ref{ed}) into Eq.(\ref{dddd}) we get for
$m=2\tilde{Z}_{kp}(0)e_ke_p$ we have:
\begin{eqnarray}\nonumber
&&(-i\partial_x Z_{kp}(-x))g_p(x)e_k+(-i\partial_x Z_{kp}(x))g_k(x)e_p\\\nonumber&&-tZ_{kp}(-x)e_p\delta(x) e_k\\&&-tZ_{kp}(x)e_k\delta(x) e_p+2t\tilde{Z}_{kp}(0)e_k e_p=0\\\label{cond0}\nonumber
&&2\tilde{Z}_{kp}(0)e_k e_p\\&&=\frac{2t(Z_{kp}(0)g_p(0)e_k+Z_{kp}(0)g_k(0)e_p)}{p+k-U-2\epsilon_d}\label{Z^*}
\end{eqnarray}
Now take $Z_{kp}(x)=e^{-i\phi_{kp}}\theta(-x)+e^{i\phi_{kp}}\theta(x)$ we get $\tan(\phi_{kp})=\frac{-Ut^2}{(k-p)(p+k-U-2\epsilon_d)}$ and $\tilde{Z}_{kp}(0)=\frac{k+p-2\epsilon_d}{k+p-U-2\epsilon_d}Z_{kp}(0)$. Define $\Gamma\equiv \frac{t^2}{2}$ and $B(k)\equiv k(k-2\epsilon_d-U)$ as in Ref.~\onlinecite{Kawakami} we can rewrite $\tan(\phi_{kp})=\frac{-2U\Gamma}{(B(k)-B(p))}$.\\

From Eq.(\ref{oo}) we can write $h(x_1,x_2)$ as:
\begin{eqnarray}\nonumber
h(x_1,x_2)&=&Z^{oo}_{kp}(x_1-x_2)h_k(x_1)h_p(x_2)\\&+&Z^{oo}_{kp}(x_2-x_1)h_k(x_2)h_p(x_1)\label{h_oo}
\end{eqnarray}
with arbitrary $Z^{oo}_{kp}(x_1-x_2)$. Now write $j(x_1,x_2)$ as:
\begin{eqnarray}\nonumber
j(x_1,x_2)&=&Z^{eo}_{kp}(x_1-x_2)g_k(x_1)h_p(x_2)\\&+&Z^{eo}_{kp}(x_2-x_1)h_k(x_2)g_p(x_1)\label{j_eo}
\end{eqnarray}
again with $Z^{eo}_{kp}(x_1-x_2)$ undetermined. Plug Eq.(\ref{j_eo}) into Eq.(\ref{eo}) we get $o(x)$ is written as:
\begin{eqnarray}
o(x)=Z^{eo}_{kp}(-x)h_p(x)e_k+Z^{eo}_{kp}(x)h_k(x)e_p\label{o_od}
\end{eqnarray}
Now if we choose $Z^{eo}_{kp}(x_1-x_2)=Z_{kp}(x_1-x_2)$ and plug Eq.(\ref{j_eo}) and Eq.(\ref{o_od}) into Eq.(\ref{od}) we get:
\begin{eqnarray}\nonumber
&&(-k+\epsilon_d)Z_{kp}(-x)h_p(x)e_k+(-p+\epsilon_d)Z_{kp}(x)h_k(x)e_p
\\\nonumber&&+t(Z_{kp}(-x)h_p(x)g_k(0)+Z_{kp}(x)h_k(x)g_p(0))\\\nonumber&&+(-i)(\partial_xZ_{kp}(-x))h_p(x)e_k+
(-i)(\partial_xZ_{kp}(x))h_k(x)e_p\\
&&=-2\sin(\phi_{kp})(h_p(0)e_k-h_k(0)e_p)=0\label{cond}
\end{eqnarray}
To satisfy Eq.(\ref{cond}) we can set $h_p(0)=0$ for arbitrary $p$. This can be done by choosing $s_{op}=-4$ in Eq.(\ref{single}). Now since $Z^{oo}_{kp}(x_1-x_2)$ is arbitrary we can choose $Z^{oo}_{kp}(x_1-x_2)=Z_{kp}(x_1-x_2)$. Also from Eq.(\ref{Z^*}) we have
\begin{eqnarray}
\tilde{Z}_{kp}(0)=\frac{p+k-2\epsilon_d}{p+k-U-2\epsilon_d}Z_{kp}(0)
\end{eqnarray}
Since the Hamiltonian in Eq.(\ref{UAn}) has rotational invariance the general form of scattering matrix for particles with momentum $k,p$ and spins $a_1,a_2$ is given by:
\begin{eqnarray}
S_{a_{1}a_{2}}^{a_{1}^{'}a_{2}^{'}}(k,p)=b(k,p)+c(k,p)\hat{P}_{a_{1}a_{2}}^{a_{1}^{'}a_{2}^{'}}\label{Smat}
\end{eqnarray}
where $\hat{P}_{a_{1}a_{2}}^{a_{1}^{'}a_{2}^{'}}=\frac{1}{2}(1_{a_1}^{a_{1}^{'}}\cdot 1_{a_2}^{a_{2}^{'}}+\vec{\sigma}_{a_1}^{a_{1}^{'}} \cdot \vec{\sigma}_{a_2}^{a_{2}^{'}})$ is the permutation operator in spins. For antiparallel spins (singlet state) as shown above $\hat{P}_{a_{1}a_{2}}^{a_{1}^{'}a_{2}^{'}}=-1$ thus we have:
\begin{eqnarray}\nonumber
b(k,p)-c(k,p)&=&\frac{Z_{kp}(x>0)}{Z_{kp}(x<0)}\\&=&\frac{B(k)-B(p)-i2U\Gamma}{B(k)-B(p)+i2U\Gamma}\label{singlet}
\end{eqnarray}
For the triplet state ($\hat{P}_{a_{1}a_{2}}^{a_{1}^{'}a_{2}^{'}}=1$) the interaction term with the impurity is absent and the particles passing through each other without changing their phase
\begin{eqnarray}
b(k,p)+c(k,p)=1\label{triplet}
\end{eqnarray}
Thus from Eq.(\ref{singlet}) and Eq.(\ref{triplet}) we get the two particle S-matrix as:
\begin{eqnarray}
\hat{S}(k,p)_{a_{i}a_{j}}^{a_{i}^{'}a_{j}^{'}}=\frac{(B(k)-B(p))\bm {\mathrm{I}}_{a_{i}a_{j}}^{a_{i}^{'}a_{j}^{'}}+i2U\Gamma\bm{\mathrm{P}}_{a_{i}a_{j}}^{a_{i}^{'}a_{j}^{'}}}{B(k)-B(p)+i2U\Gamma}\label{SmatU1}
\end{eqnarray}
Thus the integrability of two lead with Anderson type dot system is the similar to the integrability of one lead Anderson model.

 The choice of identical two particles S-matrices (by choosing $s_{op}=-4$) enables us to construct the scattering state labeled by lead indices by choosing appropriate $A,B,C$ in this even-odd basis. For example, if both particles are coming from lead 1, we shall choose $(A,B,C)=A_0(\frac{t^2}{t_2^2},\frac{-t^2}{t_1t_2},\frac{t^2}{t_1^2})$ such that the amplitude of incoming state from lead 2 is zero ($A_0$ being an overall renormalization constant). We can therefore label the eigenstate by the incoming state from lead $i$ and/or lead $j$. Without this $s_{op}$ term we cannot write
back from even-odd basis to lead indices basis in this two leads Anderson model and similarly in IRLM in Ref.~\onlinecite{Mehta}.

\section{Equivalence of TBA and SBA in equilibrium}\label{B}
Eq.(\ref{dd}) can be proved to be exact by comparing with the traditional Bethe Ansatz where
	$\langle\sum_\sigma d_\sigma^\dagger d_\sigma\rangle=
	2\int_{B}^{\infty} \dd\lambda\,\sigma_{\mathrm{imp}}(\lambda)$
	with impurity density $\sigma_{\mathrm{imp}}(\lambda)$ given by
	\begin{equation}
	\label{ddT}
	\sigma_{\mathrm{imp}}(\lambda)=\frac{\delta_{p^+}+\delta_{p^-}}{2\pi}
	-\int_B^{\infty}\dd\lambda'\, K(\lambda-\lambda')\sigma_{\mathrm{imp}}(\lambda')
	\end{equation}
	The driving term (first term) of Eq.(\ref{ddT}) is expressed by bare phase shift $\delta_{p^+}+\delta_{p^-}$
  and thus we can view $\sigma_{\mathrm{imp}}(\lambda)$ as the dressed phase shift across the impurity.	
	By comparing Eq.(\ref{ddT}) and Eq.(\ref{bulk}) in equilibrium
	($\sigma_{i}(\lambda)=\sigma_b(\lambda)$ describing bulk quasi-particle density when $B_1=B_2=B$.)
	we get
	\begin{multline}
	\label{comtba}
	\int_{B}^{\infty}\dd\lambda\,\sigma_{\mathrm{imp}}(\lambda)
	\left(\frac{-1}{\pi}\frac{\dd x(\lambda)}{\dd\lambda}\right)=\\
	2\int_{B}^{\infty}\dd\lambda\,\sigma_b(\lambda)\left(\frac{\delta_{p^+}+\delta_{p^-}}{2\pi}\right)
	\end{multline}
	by noting that the integration kernel $K(\lambda-\lambda')$ is symmetric in $\lambda$ and $\lambda'$.
	Since the equality is true for arbitrary $B$ we can also rewrite Eq.(\ref{comtba}) as
	\begin{eqnarray}
	\nonumber
	\int_{B}^{\infty}\dd\lambda\,\sigma_{\mathrm{imp}}(\lambda)&=&
	2\int_{B}^{\infty}\dd\lambda\, \sigma_b(\lambda)\left(\frac{\delta_{p^+}+\delta_{p^-}}
	{-2\frac{\dd x(\lambda)}
	{\dd\lambda}}\right)
	\\\nonumber
	&\equiv&2\int_{B}^{\infty}\dd\lambda\, \sigma_b(\lambda)\nu^{TBA}(\lambda)
	\end{eqnarray}
 	and the resulting $\nu^{TBA}(\lambda)$ is given by
	\begin{multline}
	\nu^{TBA}(\lambda)=\frac{-\tilde{x}(\lambda)\frac{y'(\lambda)}{x'(\lambda)}
	-\tilde{y}_-(\lambda)}{\tilde{x}^2(\lambda)+\tilde{y}_+^2(\lambda)}\\
	+\frac{\tilde{x}(\lambda)\frac{y'(\lambda)}{x'(\lambda)}
	+\tilde{y}_+(\lambda)}{\tilde{x}^2(\lambda)+\tilde{y}_+^2(\lambda)}
	\end{multline}
Now let us show the computation for $\nu^{SBA}(\lambda)$. First we
write one particle state of Eq.(\ref{UAn}) in even channel (with $s_{ek}=0$ for the moment) as
\begin{eqnarray}
&& |k,\sigma\rangle=\int e^{ikx}\alpha^\dagger_{ek,\sigma}(x)dx|0\rangle\\\nonumber &&=\int e^{ikx}\{(\bar{\theta}+A_k\theta)\psi_{e\sigma}^\dagger+B_k d_\sigma^\dagger\delta(x)\}dx|0\rangle
\end{eqnarray}
 Solving $\hat{H}|k,\sigma\rangle=k |k,\sigma\rangle$ we get
\begin{eqnarray}\nonumber
&&-i(-1+A_k)+B_k t=0\\\nonumber
&&\epsilon_d B_k+t\frac{1+A_k}{2}=k B_k
\end{eqnarray}
Thus we get $A_k=\frac{k-\epsilon_d-i\frac{t^2}{2}}{k-\epsilon_d+i\frac{t^2}{2}}$ and  $B_k=\frac{t}{k-\epsilon_d+i\frac{t^2}{2}}$. We may also define $g_k(x)=e^{ipx}(\bar{\theta}+A_k\theta)$ and $e_k=B_k$ to have easier comparison with Wiegmann and Tsvelick's work\cite{PB}.
The two particles state is obtained by constructing
product of two $\alpha^\dagger_{ep,\sigma}(x)$ particles state with appropriate two particles S-matrix expressed in $Z_{k^+k^-}(x_1-x_2)$.

In principle we shall use $|\Psi,N_1,N_2\rangle$ as the many body state to compute expectation value. However the simplification here, similar to the case of
IRLM in Ref.\onlinecite{Mehta}, is that different $\lambda$ (corresponding to different $p(\lambda)$) are orthogonal to each other in $L\rightarrow\infty$ limit.
Thus the many body expectation value can be obtained via two body computation and the rest just get canceled by normalization factor. To put it more explicitly let us denote $p_i$ as the real part of the complex pair $p_i^{\pm}$. Different $|p_i\rangle$ is orthogonal to each other under the condition of size of the leads taken to infinity, or $\frac{\langle p_i|p_j\rangle}{\langle p_i|p_i\rangle}\rightarrow 0$ as $L\rightarrow\infty$ for $i\neq j$. Thus the evaluation of matrix element for operator $\hat{o}$ is given by $$\frac{\langle p_1,p_2,\ldots|\hat{o}|p_1,p_2,\ldots\rangle}{\langle p_1,p_2,\ldots|p_1,p_2,\ldots\rangle}=\sum_{p_i}\frac{\langle p_i|\hat{o}|p_i\rangle}{\langle p_i|p_i\rangle}$$. Based on this result we demonstrate the
explicit computation for dot occupation by two particles wavefunctions in the following.

Denote $|\Psi\rangle$ as the two particles solution. We may write spin singlet state as
\begin{widetext}
\begin{eqnarray}\nonumber
|\Psi\rangle&=&\int d x_1 d x_2 \mathcal{A}\left\{e^{i(kx_1+px_2)}Z_{kp}(x_1-x_2)
\alpha^\dagger_{ek,\uparrow}(x_1)\alpha^\dagger_{ep,\downarrow}(x_2)\right\}|0\rangle\\\nonumber
&=&\int dx_1 d x_2 \Big\{Z_{kp}(x_1-x_2)\{g_k(x_1)g_p(x_2)\psi_\uparrow^\dagger(x_1)\psi_{e\downarrow}^\dagger(x_2)+g_k(x_1)e_p\psi_\uparrow^\dagger(x_1)d_\downarrow^\dagger\delta(x_2)\\\nonumber
&&+\, e_k g_p(x_2)d_\uparrow^\dagger\delta(x_1)\psi_\downarrow^\dagger(x_2)+e_k e_pd_\uparrow^\dagger d_\downarrow^\dagger\delta(x_1)\delta(x_2)\}-Z_{kp}(x_2-x_1)\{g_k(x_2)g_p(x_1)
\psi_{e\downarrow}^\dagger(x_2)\psi_{e\uparrow}^\dagger(x_1)\\\nonumber
&&+\, g_k(x_2)e_p\psi_{e\downarrow}^\dagger(x_2)d_\uparrow^\dagger\delta(x_1)
+\,e_k g_p(x_1)d_\downarrow^\dagger\delta(x_2)\psi_{e\uparrow}^\dagger(x_1)+e_k e_pd_\downarrow^\dagger d_\uparrow^\dagger\delta(x_1)\delta(x_2)\}\Big\}|0\rangle\\\nonumber
&=&\Big\{\int d x_1 d x_2 [Z_{kp}(x_1-x_2)g_k(x_1)g_p(x_2)+Z_{kp}(x_2-x_1)g_k(x_2)g_p(x_1)]\psi^\dagger_{e\uparrow}(x_1)\psi^\dagger_{e\downarrow}(x_2)\\\nonumber
&&+\int d x [Z_{kp}(x)g_k(x)e_p+Z_{kp}(-x)g_p(x)e_k](\psi_{e\uparrow}^\dagger(x)d^\dagger_\downarrow-\psi_{e\downarrow}^\dagger(x)d^\dagger_\uparrow)
+ 2e_k e_p \tilde{Z}_{kp}(0)d_\downarrow^\dagger d_\uparrow^\dagger\Big\}|0\rangle
\label{2pwf}
\end{eqnarray}
\end{widetext}
With $\mathcal{A}$ denoting anti-symmetrization and $\tilde{Z}_{kp}(0)=\frac{k+p-2\epsilon_d}{k+p-U-2\epsilon_d}Z_{kp}(0)$.

Solving $\hat{H}|k,\sigma;p,-\sigma\rangle=(k+p)|k,\sigma;p,-\sigma\rangle$ we obtain
\begin{widetext}
\begin{equation}\nonumber
Z_{kp}(x_1-x_2)=\theta(x_1-x_2)+\frac{(k-p)(k+p-2\epsilon_d - U) - i Ut^2}{(k-p)(k+p-2\epsilon_d - U)+ i Ut^2}\theta(x_2-x_1)
\end{equation}
\end{widetext}
For the case of bound state the two particle S-matrix is given by $Z_{k^+k^-}(x_1-x_2)=\theta(x_1-x_2)\equiv\theta^x_{12}$. The normalization factor and matrix element of dot occupation given by the even channel two particles wavefunction are
\begin{widetext}
\begin{eqnarray}
\nonumber
\langle\Psi|\Psi\rangle&=&\int dy_1dy_2\int dx_1dx_2 (\theta^y_{12}g_{k^+}(y_1)g_{k^-}(y_2)+\theta^y_{21}g_{k^+}(y_2)g_{k^-}(y_1))^\ast\\\nonumber
&&\times(\theta^x_{12}g_{k^+}(x_1)g_{k^-}(x_2)+\theta^x_{21}g_{k^+}(x_2)g_{k^-}(x_1))\delta(x_1-y_1)\delta(x_2-y_2)\\\nonumber
&&+2\int dy\int dx [\theta(y)g_{k^+}(y)e_{k^-}+\theta(-y)g_{k^-}(y)e_{k^+}]^\ast[\theta(x)g_{k^+}(x)e_{k^-}+\theta(-x)g_{k^-}(x)e_{k^+}]\delta(x-y)\\\nonumber
&&+4(e_{k^+} e_{k^-} \tilde{Z}_{k^+k^-}(0))^\ast(e_{k^+} e_{k^-} \tilde{Z}_{k^+k^-}(0))\\\nonumber
\sum_{\sigma}\langle\Psi| \hat{d}^\dagger_\sigma\hat{d}_\sigma|\Psi\rangle &=&
2\int dy\int dx [\theta(y)g_{k^+}(y)e_{k^-}+\theta(-y)g_{k^-}(y)e_{k^+}]^\ast[\theta(x)g_{k^+}(x)e_{k^-}+\theta(-x)g_{k^-}(x)e_{k^+}]\delta(x-y)\\\nonumber
&&+8(e_{k^+} e_{k^-} \tilde{Z}_{k^+k^-}(0))^\ast(e_{k^+} e_{k^-} \tilde{Z}_{k^+k^-}(0))\\\nonumber
&=&2\left\{\int dx [\theta(x)|g_{k^+}(x)e_{k^-}|^2+\theta(-x)|g_{k^-}(x)e_{k^+}|^2]+4|e_{k^+} e_{k^-} \tilde{Z}_{k^+k^-}(0)|^2\right\}
\end{eqnarray}
\end{widetext}
Note that the even channel bound state can be written as sum over bound state of $\{11,12,21,22\}$ ($4$ strings type) or $\{11,22\}$ ($2$ strings type)
with the same real part of energy $k=x(\lambda)$. This can be viewed as the consistency counting from Fock basis to Bethe basis as electrons in lead 1 and lead 2
has $4$ fold degeneracies in its initial state ($2$ different spins in each lead).
Also note that
\begin{widetext}
\begin{eqnarray}\nonumber
\int dx_1 dx_2\ \theta^x_{12}|g_{k^+}(x_1)g_{k^-}(x_2)|^2
&=&\int dx_1 dx_2 |e^{i(k^+x_1+k^-x_2)}(\bar{\theta}_1+\theta_1A_{k^+})(\bar{\theta}_2+\theta_2A_{k^-}) |^2\theta_{12}\\\nonumber
&=&\int dx_1 dx_2\ e^{-2\xi_k(x_1-x_2)}|\bar{\theta}_1\bar{\theta}_2\theta_{12}+\theta_1\bar{\theta}_2\theta_{12}A_{k^+}+\theta_1\theta_2\theta_{12}A_{k^+}A_{k^-}|^2\\\nonumber
&=&\left(\frac{L}{2\xi_k}-\frac{1-e^{-2\xi_k L}}{(2\xi_k)^2}\right)\left(1+|A_{k^+}A_{k^-}|^2\right)
+\left(\frac{1-e^{-2\xi_k L}}{2\xi_k}\right)^2|A_{k^+}|^2\\\nonumber
\int dx\, \theta(x)|g_{k^+}(x)e_{k^-}|^2&=&\int dx\, \theta(x)|e^{i(k+i\xi_k)x}(\theta(-x)+A_{k^+}\theta(x))e_{k^-}|^2=\int_0^L dx\, e^{-2\xi_k x}|A_{k^+}e_{k^-}|^2\\\nonumber
&=&\frac{1}{2\xi_k}\Bigg|\frac{k-\epsilon_d+i\xi_k-i\Gamma}{k-\epsilon_d+i\xi_k+i\Gamma}
\frac{t}{k-\epsilon_d-i\xi_k+i\Gamma}\Bigg|^2
=\frac{1}{2\xi_k}\Bigg|\frac{t}{k-\epsilon_d+i\xi_k+i\Gamma}\Bigg|^2\\\nonumber
\int dx\, \theta(-x)|g_{k^-}(x)e_{k^+}|^2
&=&\int dx\, \theta(-x)|e^{i(k-i\xi_k)x}(\theta(-x)+A_{k^-}\theta(x))e_{k^+}|^2
=\int_{-L}^0dx\, e^{2\xi_k x}|A_{k^-}e_{k^+}|^2\\\nonumber
&=&\frac{1}{2\xi_k}\Bigg|\frac{t}{k-\epsilon_d+i\xi_k+i\Gamma}\Bigg|^2
\end{eqnarray}
\end{widetext}
with $\tilde{Z}_{k^+k^-}(0)=\frac{2(k-\epsilon_d)}{2(k-\epsilon_d)-U}Z_{k^+k^-}(0)$ and $Z_{k^+k^-}(0)=\frac{1}{2}$ based on our regularization scheme. By expressing $k=x(\lambda)$ and $\xi_k=y(\lambda)$ and taking $L\rightarrow\infty$ thus preserving $\frac{1}{L}$ terms only we get
\begin{widetext}
\begin{eqnarray}
\label{dd2}
\frac{\langle\Psi|\sum_{\sigma}\hat{d}_\sigma^\dagger\hat{d}_\sigma|\Psi\rangle}{\langle\Psi|\Psi\rangle}&=&\frac{1}{L}\nu^{SBA}(\lambda)\\\nonumber
&=&\frac{1}{L}\left\{\frac{2\Gamma}{\tilde{x}^2(\lambda)+\tilde{y}^2_+(\lambda)}+
\frac{16y(\lambda)\Gamma^2}{(\tilde{x}^2(\lambda)+\tilde{y}^2_-(\lambda))(\tilde{x}^2(\lambda)+\tilde{y}^2_+(\lambda))}\left(\frac{\tilde{x}(\lambda)}{2\tilde{x}(\lambda)-U}\right)^2\right\}\ .
\end{eqnarray}
\end{widetext}
	By expressing $\nu^{TBA}(\lambda)$ and $\nu^{SBA}(\lambda)$ in $\lambda$ explicitly
	we see that $\nu^{TBA}(\lambda)=\nu^{SBA}(\lambda)$.
	Since $\langle\sum_\sigma d^\dagger_\sigma d_\sigma\rangle=2\int_B^{\infty}d\lambda\sigma_{imp}(\lambda)$
	in TBA we have proved that the expectation value evaluated by the
	state we constructed is exact and the equivalence of SBA and TBA in equilibrium in this two-lead Anderson model.


\begin{thebibliography}{999}
\bibitem{footnote3} This dressed energy is the sum of the dressed energy of spinon and that of antiholon nearby equilibrium Fermi surface.
\bibitem{footnote4} The equality can be proved analytically in linear response. In numerics we also see very good agreement, especially for large voltage, with voltage computed by $\mu_1-\mu_2$ given by free energy and voltage computed by difference in particle number.
\end{thebibliography}
\end{document}